\documentclass[a4paper,11pt]{article}
\pdfoutput=1 

\usepackage{jinstpub} 
\usepackage{multirow}
\usepackage{graphicx}
\usepackage{chemfig,siunitx}
\usepackage{epstopdf}
\usepackage{amsmath}
\usepackage{mathrsfs}
\usepackage{amsfonts}
\usepackage{amssymb}
\usepackage{textcomp}
\usepackage{subfig}
\usepackage{enumitem}
\usepackage{multirow}
\usepackage{enumitem}
\usepackage{ulem} 
\usepackage[sort&compress]{natbib}

\newcommand{\Nucl}[2]{$\mathrm{^{#2}#1}$}
\newcommand{\ag}{$\alpha$}

\long\def\/*#1*/{}

\title{First detection of radon progeny recoil tracks by MIMAC} 

\author[a]{Q.~Riffard}
\author[a]{D.~Santos}
\author[a]{O.~Guillaudin}
\author[a]{G.~Bosson}
\author[a]{O.~Bourrion}
\author[a]{J.~Bouvier}
\author[a]{T.~Descombes}
\author[a]{C.~Fourel}
\author[a]{J.-F.~Muraz}
\author[b]{L.~Lebreton}
\author[b]{D.~Maire}
\author[b]{P.~Colas}
\author[c]{E.~Ferrer-Ribas}
\author[c]{I.~Giomataris}
\author[d]{J.~Busto}
\author[d]{D.~Fouchez}
\author[d]{J.~Brunner}
\author[d,e]{C.~Tao}

\affiliation[a]{LPSC, Universit\'e Grenoble-Alpes, CNRS/IN2P3, Grenoble, France}
\affiliation[b]{LMDN, IRSN Cadarache, 13115 Saint-Paul Lez-Durance, France}
\affiliation[c]{IRFU, CEA Saclay, 91191 Gif-sur-Yvette, France}
\affiliation[d]{Aix Marseille Universit\'e, CNRS/IN2P3, CPPM UMR 7346, 13288, Marseille, France}
\affiliation[e]{Tsinghua Center for Astrophysics, Tsinghua University, Beijing 100084, China}

\emailAdd{riffard@apc.in2p3.fr}
\emailAdd{santos@lpsc.in2p3.fr}

\abstract{
	The MIMAC experiment is a \textmu-TPC project for directional dark matter search. 
Directional detection strategy is based on the measurement of the WIMP flux anisotropy due to the solar system motion with respect to the dark matter halo.
The main purpose of MIMAC project is the measurement of nuclear recoil energy and 3D direction from the WIMP elastic scattering on target nuclei. Since June 2012 a bi-chamber prototype is operating at the Modane underground laboratory.
In this paper, we report the first ionization energy and 3D track observations of NRs produced by the radon progeny. This measurement shows the capability of the MIMAC detector and opens the possibility to explore the low energy recoil directionality signature. }

\arxivnumber{1504.05865}


\begin{document}

\maketitle 
\flushbottom


\section*{Introduction}

A large number of astrophysical and cosmological observations at different scales support the existence of a cold dark matter component in the Universe. At the Universe scale, this component represents roughly 26\% of the total mass-energy density of the Universe~\cite{PlanckCollaboration2015a}. The Weakly Interacting Massive Particle (WIMP), a generic particle, is one of the leading dark matter particle candidates: a massive particle interacting only through weak and gravitational interactions. This candidate is supported by the so-called "WIMP miracle" - the fact that such a particle, produced in the early Universe, would give the correct dark matter abundance - and naturally occurs in R-parity conserving supersymmetric models.
At the Milky Way scale, dark matter forms a static halo surrounding our galaxy. 
The relative motion of the solar system through the dark halo produces a flux of WIMP on Earth.
Through the weak interaction, WIMP could interact with ordinary matter producing nuclear recoils (NR) by elastic scattering~\cite{Goodman_1984dc}. 
In the last two decades, a large experimental effort has been deployed by international collaborations in order to probe a direct detection of NR from WIMP-nucleus interactions without  success~\cite{Agnes2016,Akerib2016a,XENON100Collaboration2016a}. The main goal of these experiments is to improve their sensitivity by increasing the exposure and/or by reducing the energy threshold.
The major limitation of the direct search strategy arises from an irreducible background: the neutrino-induced NR. Indeed, neutrinos can produce NRs with a kinetic energy at the 10~keV scale by coherent scattering limiting the detector sensitivity~\cite{Billard:2013qya}. 

The directional detection strategy, first proposed in 1988~\cite{Spergel1988} is based on the fact that the WIMP event distribution is expected to have an excess in the direction of the Solar system motion with respect to the galactic center. It induces a dipole feature in the WIMP-produced NR distribution~\cite{Billard:2009mf}, whereas the cosmic-ray and $(\alpha,n)$ induced background distributions~\cite{Mei:2005gm} are expected to be isotropic in the galactic rest frame or completely uncorrelated with respect to the motion around the galactic center. Several directional features from different backgrounds would display unambiguous differences with the WIMP signal: dipole~\cite{Billard:2009mf}, ringlike~\cite{Bozorgnia:2011vc} and aberrations~\cite{Bozorgnia:2011tk}. They may be used to either exclude dark matter~\cite{Billard:2010gp,Henderson:2008bn}, discover galactic
dark matter with a high significance~\cite{Billard:2009mf,Billard:2011zj,Green:2010zm} or constrain WIMP and halo properties~\cite{Billard2010,Alves:2012ay,Lee:2014cpa,O'Hare:2014oxa}, depending on the WIMP-nucleon cross section. 
It has also been recognized as the ultimate detection strategy to look for dark matter beyond the neutrino floor~\cite{Billard:2013qya,Grothaus:2014hja,Ruppin:2014bra}.

The MIMAC experiment is a directional detection project initiated in 2007 at the Laboratoire de Physique Subatomique et de Cosmologie (LPSC Grenoble-France)~\cite{Santos2007} in collaboration with IRFU-Saclay, CPPM- Marseille, IRSN and recently the Tsinghua University. As the other directional detection experiments~\cite{Ahlen2009,Battat_2016pap}, the aim of the MIMAC project is the measurement of the WIMP-induced NR energy and angular distributions in order to constrain the WIMP properties. 
In this context, the MIMAC collaboration developed an original detector and readout strategy allowing the measurement of NR tracks in 3D.
Since June 2012, a bi-chamber prototype was installed at the Modane underground laboratory (Laboratoire Souterrain de Modane\footnote{LSM}) for preliminary tests.
This first data taking points the radon progeny as the current most important source of background for the MIMAC experiment.
In this paper the first 3D tracks measurement of the daughter NRs from radon progeny \ag-decay is presented.


\begin{figure}[tbp]

\centering
		\includegraphics[width=1\linewidth]{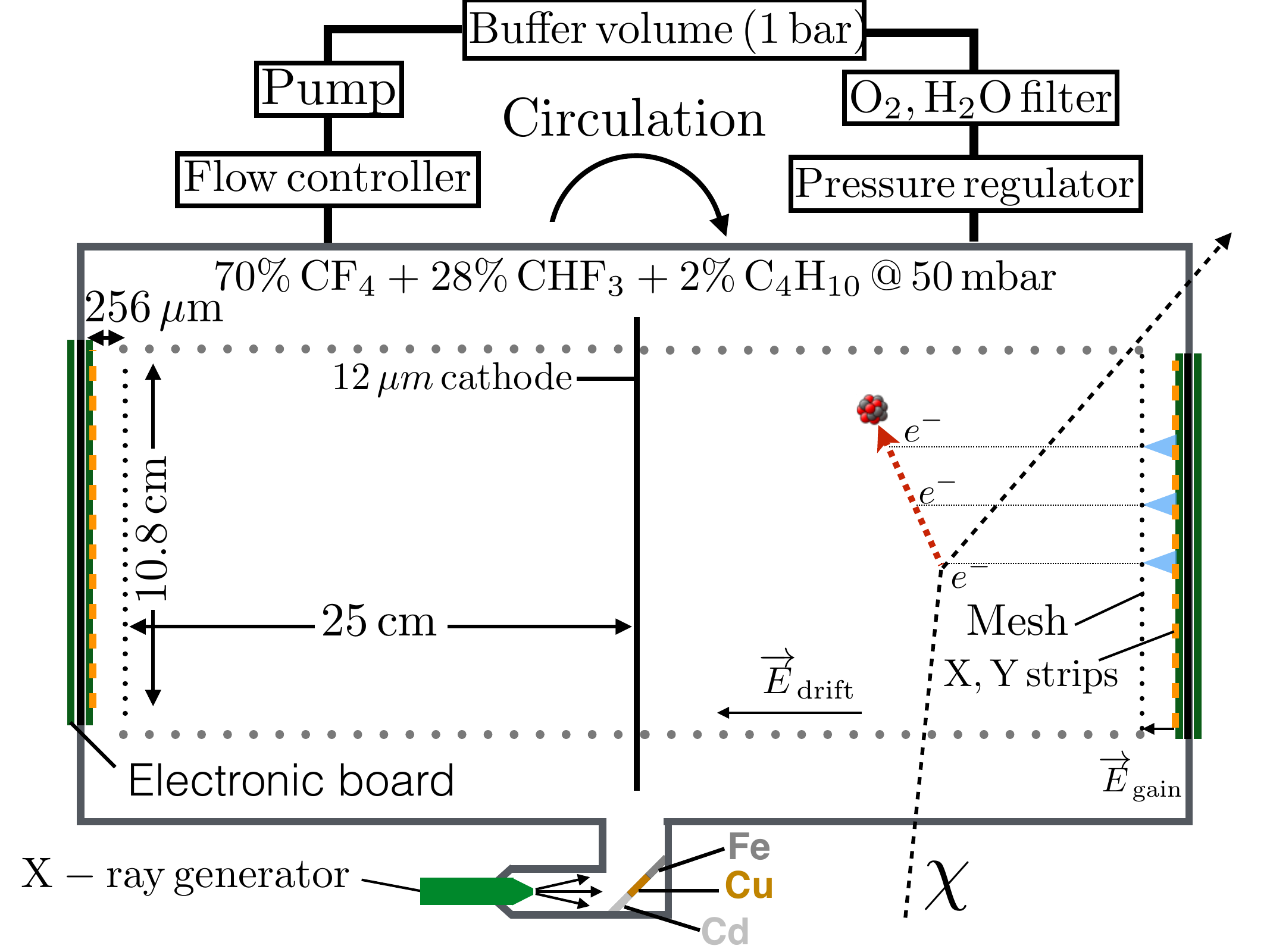}
		\includegraphics[width=0.65\linewidth]{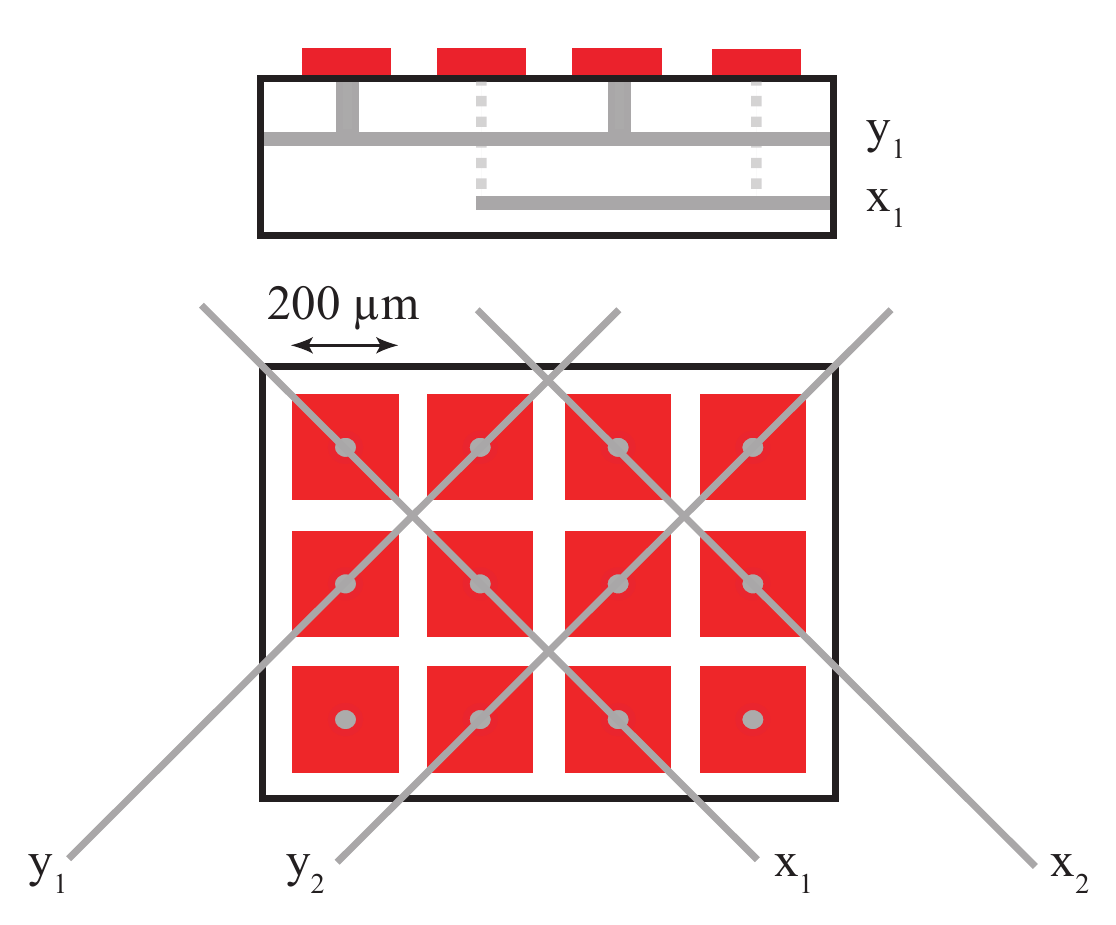}
		\caption{ Top panel: Illustration of the bi-chamber prototype configuration and the ionization electron collection from a NR produced by the WIMP elastic scattering (not~in~scale). 
				Bottom panel: Scheme of the X and Y strips arrangement of the Micromegas. Each pixel is 200\textmu m wide with a 100\textmu m gap corresponding to a 424\textmu m pitch.}
		\label{fig:Bichambre}
\end{figure}

\section{The MIMAC Experiment}
\label{sec:MIMACExp}

The main goal of the MIMAC experiment is to define the large gaseous TPC for directional detection of WIMP~\cite{Santos2007} by using a $\mathrm{CF_4}$-based low pressure (at $50\,\mathrm{mbar}$) gas mixture. The concept of this experiment is a replication of active cells to reach a final active volume of $50m^3$. Each cell consists of a gaseous TPC read by a pixelated Micromegas coupled with a self-triggered fast electronic system allowing the measurement of the NR energy and 3D tracks. To demonstrate the feasibility of the chosen technology, a bi-chamber prototype, containing two active cells, has been deployed at the LSM since June 2012~\cite{Santos:2013oua}. 

\subsection{The Bi-Chamber Prototype}
\label{sec:SetUp}

Top panel of figure~\ref{fig:Bichambre} represents a scheme of the bi-chamber prototype. The detector is filled with a $\mathrm{CF_4} + 28\% \mathrm{CHF_3} + 2\% \mathrm{C_4H_{10}}$ gas mixture at 50~mbar and is composed of two mirroring symmetric cells sharing a fine aluminized mylar cathode.
Each cell is equipped with a of 10.8~cm wide pixelated Micromegas~\cite{Iguaz2011,Giomataris2006} with $256\times 256$ strips on X and Y axis as represented in figure~\ref{fig:Bichambre} bottom panel.
When a WIMP (represented by a dashed line) interacts with a gas nucleus by elastic scattering, it transfers a part of its kinetic energy to the nucleus producing a NR (represented by a red dashed arrow). 
The NR releases only a part of its energy by ionization, creating electron-ion pairs. 
Ionization electrons are collected to a mesh Micromegas with a $21.4\,\mbox{\textmu m/ns}$ drift velocity (measured by using the method described in~\cite{Billard2013a}) by means of an electric field ${E_\mathrm{drift}}\sim 180\,\mathrm{V.cm^{-1}}$. Passing through the mesh these electrons are amplified by avalanche thanks to a much higher electric field ${E_\mathrm{gain}}\sim 18\,\mathrm{kV.cm^{-1}}$ (corresponding to an $\sim 470\,\mathrm{V}$ amplification voltage). 
On the one hand, the deposited ionization energy is read by a charge preamplifier connected to the mesh. Due to the saturation of the ADC our energy dynamic range is ranging from 0 to 62~keVee.
In addition, by coupling the pixelated Micromegas with a fast self-triggering electronic system~\cite{Richer2009,Bourrion2010a}, the X and Y strips are read at 50\, MHz allowing the measurement of the (X,Y) projection of the track as a function of time. Knowing the electron drift velocity, it is then possible to reconstruct the relative Z coordinate. The determination of the absolute Z coordinate of the track is not directly possible with this readout. This aspect is discussed in section~\ref{secMPD}.

This paper is focused on the analysis of the first two MIMAC data taking run in dark matter search mode. The properties of these runs are summarized on the table~\ref{tab-summRun}. The same electric fields have been used all along the data taking. We can notice a difference of total events rate between the two runs. This difference is due to a reduction of the gas contamination by radon isotopes as discussed in section~\ref{sec-pollutionEv}.

\begin{table}[h]
    \begin{center}
    \begin{tabular}{|c|cccc|}
        \hline
        Run label & Drift voltage & Gain voltage & Lived-time & Total event rate\\
        \hline
        Run 2012 & 4520 V & 470 V & 77 days & $11.2\pm 0.2\,\mathrm{min^{-1}}$\\
        Run 2013 & 4520 V & 470 V & 103 days & $3.4\pm 0.1\,\mathrm{min^{-1}}$\\
        \hline
    \end{tabular}
    \caption{Table summarizing the 2012 and 2013 data taking properties.}
    \label{tab-summRun}
    \end{center}
\end{table}

\subsection{Detector calibration}

\begin{figure}[tbp]
\centering
		\includegraphics[width=0.49\linewidth]{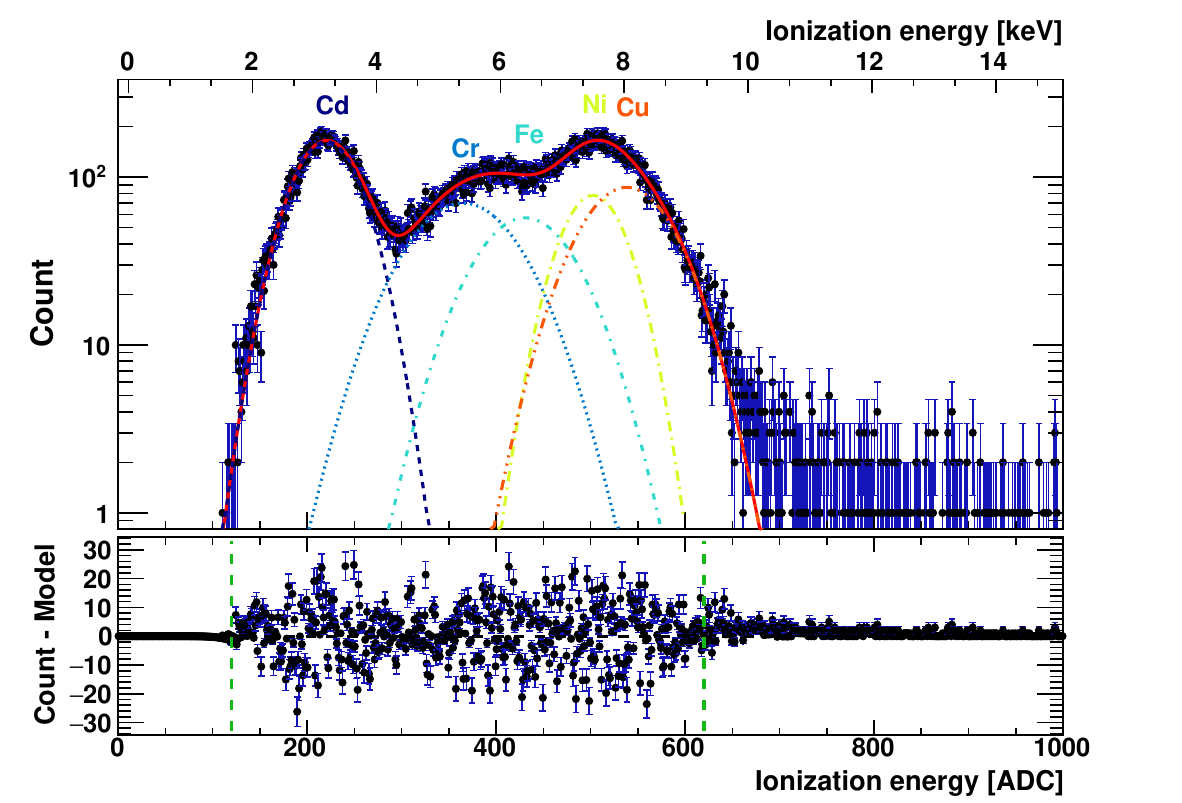}
		\includegraphics[width=0.49\linewidth]{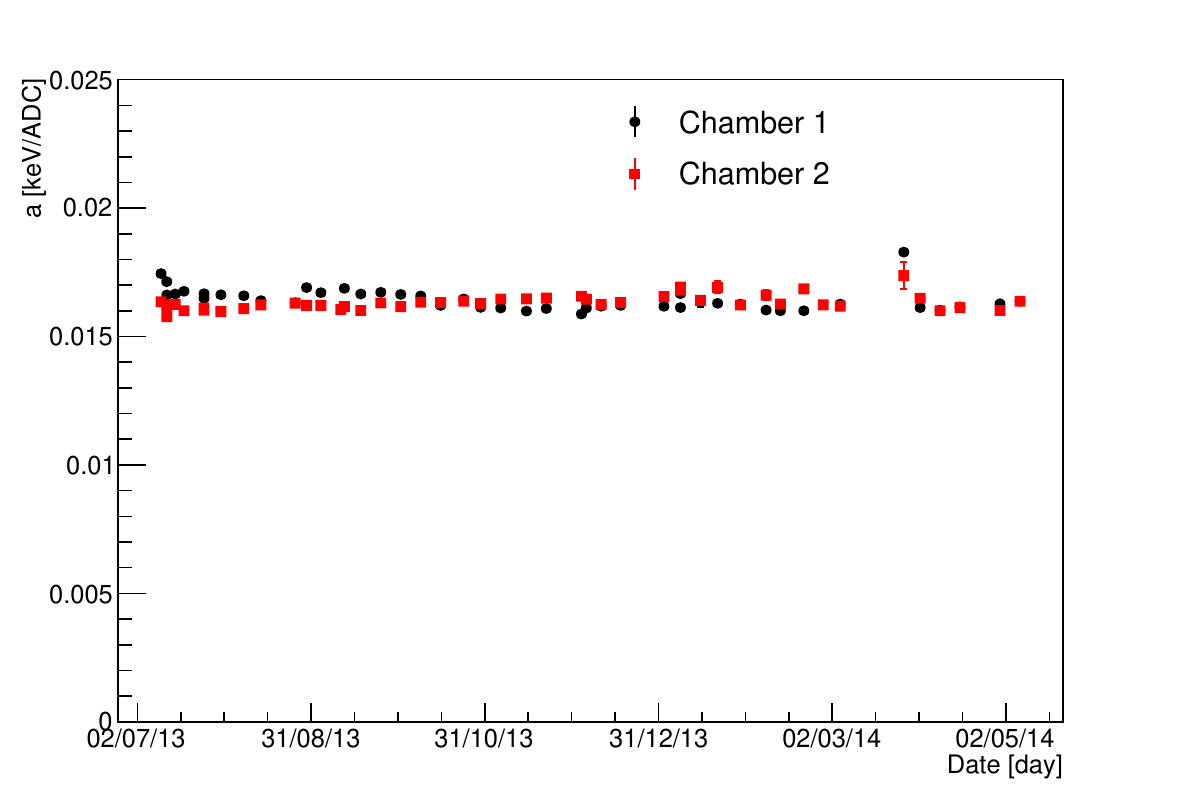}
		\caption{ Left panel: X-ray generator calibration spectrum measurement. Energy peaks (dashed lines) produced using fluorescence photons from cadmium, iron and copper metal foils. The red line represents the total fit of the spectrum. The bottom panel represents the residual of the fit. Right panel: Slope of the linear calibration $a$ as a function of time for chamber~1 (black dots) and chamber~2 (red dots).}
		\label{fig:Calibration}
\end{figure}

As shown in figure~\ref{fig:Bichambre}, an X-ray generator is permanently mounted in front of metal foils coupled to the main vessel. During irradiation, metal foils and stainless steel vessel produce fluorescence photons that can be used as calibration sources (photons from Cd, Fe, Cu, Cr and Ni).
The left panel of figure~\ref{fig:Calibration} shows a typical calibration spectrum from the chamber~1 and the best-fit (red line). We obtained an excellent agreement between model and data with a reduced $\chi^2$ of 1.09 and an associated p-value of 0.464. In addition, no structures are observable on the fit residual (bottom panel), showing that the model is fully adequate to describe the calibration spectrum. Each contribution from the metal foils is represented by dashed lines such as the 3.19~keV from Cd, the 5.4~keV from Cr, the 6.4~keV from Fe, the 7.5~keV from the Ni, the 8.04~keV from Cu fluorescence photons. 

\subsection{Gas Circulation System}

The gas circulation system developed for this experiment is an important component of the detector, ensuring the gas quality stability in a closed circuit. As schematically shown in figure ~\ref{fig:Bichambre}, it includes a buffer volume, an oxygen filter, a dry and very low leak pump and a pressure regulator. This system has been designed to prevent the presence of impurities and $\mathrm{O_2}$ into the gas. These contaminants affect the gain of the Micromegas and then the energy resolution. 
The detector is weekly calibrated in order to monitor the gain variations through the slope of the linear calibration $a$.
The right panel of figure \ref{fig:Calibration} shows the variation of $a$ as a function of time for both chambers (chamber~1 in black and chamber~2 in red). We can observe that the gains in both chambers are roughly constant with variations lower than 1\% over several months, demonstrating the gain stability during the data taking.

\section{Radon progeny recoils origin and signature}
\label{seq-Origin}

\subsection{Radon emanation origin}
\label{sec:Origins}

A background for dark matter detection arises from an intrinsic pollution of detector material by radioactive nuclei, such as \Nucl{U}{238} and \Nucl{Th}{232}. The decay chains of such nuclei  produce electron recoil (ER) background from $\beta$ decays and $\gamma$ de-excitation. In addition, we have radon emanations from the surface of the materials, releasing \Nucl{Rn}{222} and \Nucl{Rn}{220} inside the gas of the detector.
Radon Progeny Recoils\footnote{RPR} denote NRs produced by \ag-decays from \Nucl{Rn}{222} and \Nucl{Rn}{220} decay chains. It includes \ag-particles and daughter nucleus recoils. 
These events have been extensively observed in dark matter detectors, see for example:~\cite{Burgos2007,Daw:2013waa,Battat2014,Malling2013}.
The first report of RPR events in a dark matter directional detector was published by the DRIFT collaboration in 2008~\cite{Burgos2007} ~\cite{Daw:2013waa,Battat2014}. They measured  the track length of \ag-particles from the radon progeny and correlated the length of the \ag-particle track with its kinetic energy in order to identify each nucleus decay. 

When the MIMAC detector is in dark matter search mode, it is not possible to measure \ag-particle energies or their track lengths. Indeed, at 50 mbar their tracks are not fully contained in the active volume of one chamber. Instead, we can observe the daughter NRs produced by the \ag-particle emission measuring their 3D tracks with their total ionization energy showing the ability of the detector to get a clear signature of low energy NR tracks. 

\begin{table}
	\begin{center}
		\begin{tabular}{|c|c|c|c|c|c|c|c|}
			\hline
			\rule[-0.2cm]{0cm}{0.55cm} 
			Parent & $T_{1/2}$ & Mode & $E^{kin}_{\alpha/\beta\,\mathrm{max}}$ & Daughter & $E^{kin}_{recoil}$ & $E^{ioni}_{recoil}$  \\
			& & & [MeV] & & [keV] & [keVee]	\\
			
			 
      		\hline
			\hline
			\multicolumn{7}{|c|}{From $\mathrm{^{222}Rn}$} \rule[-0.15cm]{0cm}{0.5cm}\\
     	 	\hline
		\rule[0.cm]{0cm}{0.35cm}$\mathrm{^{222}Rn}$& 3.8 days & $\alpha$ & $5.489$ & $\mathrm{^{218}Po}$ & $100.8$ & $38.23$ \\
      		$\mathrm{^{218}Po}$& 3.1 min & $\alpha$ &$6.002$ & $\mathrm{^{214}Pb}$ & $112.3$ &   $43.90$ \\
		$\mathrm{^{214}Pb}$& 27 min & $\beta^-$ & 1.024  & $\mathrm{^{214}Bi}$ & -  & -   \\
		$\mathrm{^{214}Bi}$& 20 min & $\beta^-$ & 3.272 & $\mathrm{^{214}Po}$ & - & -   \\
      		$\mathrm{^{214}Po}$&164 \textmu s & $\alpha$ &$7.687$ & $\mathrm{^{210}Pb}$& $146.5$  &  $58.78$ \\
      		$\mathrm{^{210}Pb}$& 22 years & $\beta^-$ & 0.064 & $\mathrm{^{210}Bi}$ & - & -\\
		$\mathrm{^{210}Bi}$& 5 days & $\beta^-$ & 1.163 & $\mathrm{^{210}Po}$ &- &-  \\
      		$\mathrm{^{210}Po}$& 138 days & $\alpha$ & 5.304 & $\mathrm{^{206}Pb}$ (stable) & 103.7 & 40.28 \\
      		\hline
      		\hline
			\multicolumn{7}{|c|}{From $\mathrm{^{220}Rn}$} \rule[-0.15cm]{0cm}{0.5cm}\\
			\hline
      		\rule[0.cm]{0cm}{0.35cm}$\mathrm{^{220}Rn}$& 55 s& $\alpha$ & $6.288$ & $\mathrm{^{216}Po}$ & $116.5$ & $45.4$   \\
      		$\mathrm{^{216}Po}$& 0.14 s & $\alpha$ &$6.778$ & $\mathrm{^{212}Pb}$ & $128.0$ & $50.0$    \\
		$\mathrm{^{212}Pb}$& 10.6 hours & $\beta^-$ & 0.574 & $\mathrm{^{212}Bi}$ & - & -    \\
		
		$\mathrm{^{212}Bi}$ (64\%)& 61 min & $\beta^-$ & 2.254 & $\mathrm{^{212}Po}$ & - & -   \\
      	
		$\mathrm{^{212}Bi}$(36\%)& 61 min & $\alpha$ &$6.090$ & $\mathrm{^{208}Tl}$ & $117.2$ & $45.7$    \\
		$\mathrm{^{212}Po}$& 0.3 \textmu s & $\alpha$ &$8.785$ & $\mathrm{^{208}Pb}$ & $169.1$ & $69.8$    \\
		$\mathrm{^{208}Tl}$ & 3.0 min & $\beta^-$ & 5.001 & $\mathrm{^{208}Pb}$ (stable) & - & -    \\
	      	\hline
		\end{tabular}
		

		\caption{ Details of the $\alpha$ and $\beta$ decays from $\mathrm{^{222}Rn}$ and $\mathrm{^{220}Rn}$ decay chains. This table contains the half-life $T_{1/2}$ of each element and the decay mode. For $\alpha$-decays, it summarizes energies $E^{kin}_{\alpha}$ of the emitted $\alpha$-particles and the kinetic $E^{kin}_{recoil}$ and ionization $E^{ioni}_{recoil}$ energies of daughter nuclei. Ionization energies of daughter nuclei were estimated with SRIM~\cite{ZieglerJ.F.;Biersack1985}. For $\beta$-decays, it summarizes only $\beta$ maximal energies $E^{kin}_{\beta\,\mathrm{max}}$.}
		\label{tab:Progeny}
	\end{center}
\end{table}

\subsection{Radon progeny recoils event signatures}
\label{sec:Signatures}

Table~\ref{tab:Progeny} presents the considered radio nuclei from \Nucl{Rn}{222} and \Nucl{Rn}{220} decay chains.
It shows the half-life of each element and the emitted particle with their maximum kinetic energy.
From these radio-nucleus decays we can expect two different kinds of events: electrons from $\beta$ and $\gamma $ emissions, and RPR events from $\alpha$-particles and daughter nucleus emissions. 

From radon chains we can distinguish two contributions to the ER background: $\beta$ decays  and electrons produced by Compton scattering of $\gamma$-rays inside the detector. For dark matter searches, this background can be removed using a dedicated ER/NR discrimination method based on boosted decision trees (as discussed in section~\ref{sec-RPR_recoil_selection}).
 
In addition, RPR events occur at different positions inside the detector, which impacts their tracks and energies and, in consequence, their discrimination. The different positions are schematically described in figure~\ref{fig:origin}.  Red polygons represent \ag-decays, blue dots represent the daughter nuclei, plain arrows represent \ag-particles or daughter nucleus motion directions and the dashed arrows represent the daughter nucleus migration due to the drift electric field. We can distinguish five types of events as a function of their positions:
 \begin{description}[leftmargin=0.4cm]
	\item[1) Volume events ] \Nucl{Rn}{222} and \Nucl{Rn}{220} can be present everywhere in the active volume. 
	While an \ag-decay occurs in the gas volume, the energy deposited by the emitted $\alpha$-particle with a 5.5 or 6.8~MeV kinetic energy (see table \ref{tab:Progeny}) saturates the preamplifier.
	These events can be easily discriminated using a cut on the saturation energy. 
	Resultant daughters (blue dots) are in general produced with a positive electric charged. It implies that they are collected on the cathode due to the drift electric field. 
	Their drift velocity can be estimated as three orders of magnitude lower than the electron drift velocity ($v_{drift}^{e^-} = 21.4\mbox{ \textmu m}/\mathrm{ns}$ in MIMAC gas mixture~\cite{Billard2013a}) allowing daughter nuclei to reach the cathode before the next \ag-decay of the chain. At the cathode, there is an accumulation of radon progeny elements. 
				
\begin{figure}[tbp]
\centering
		\includegraphics[width=\linewidth]{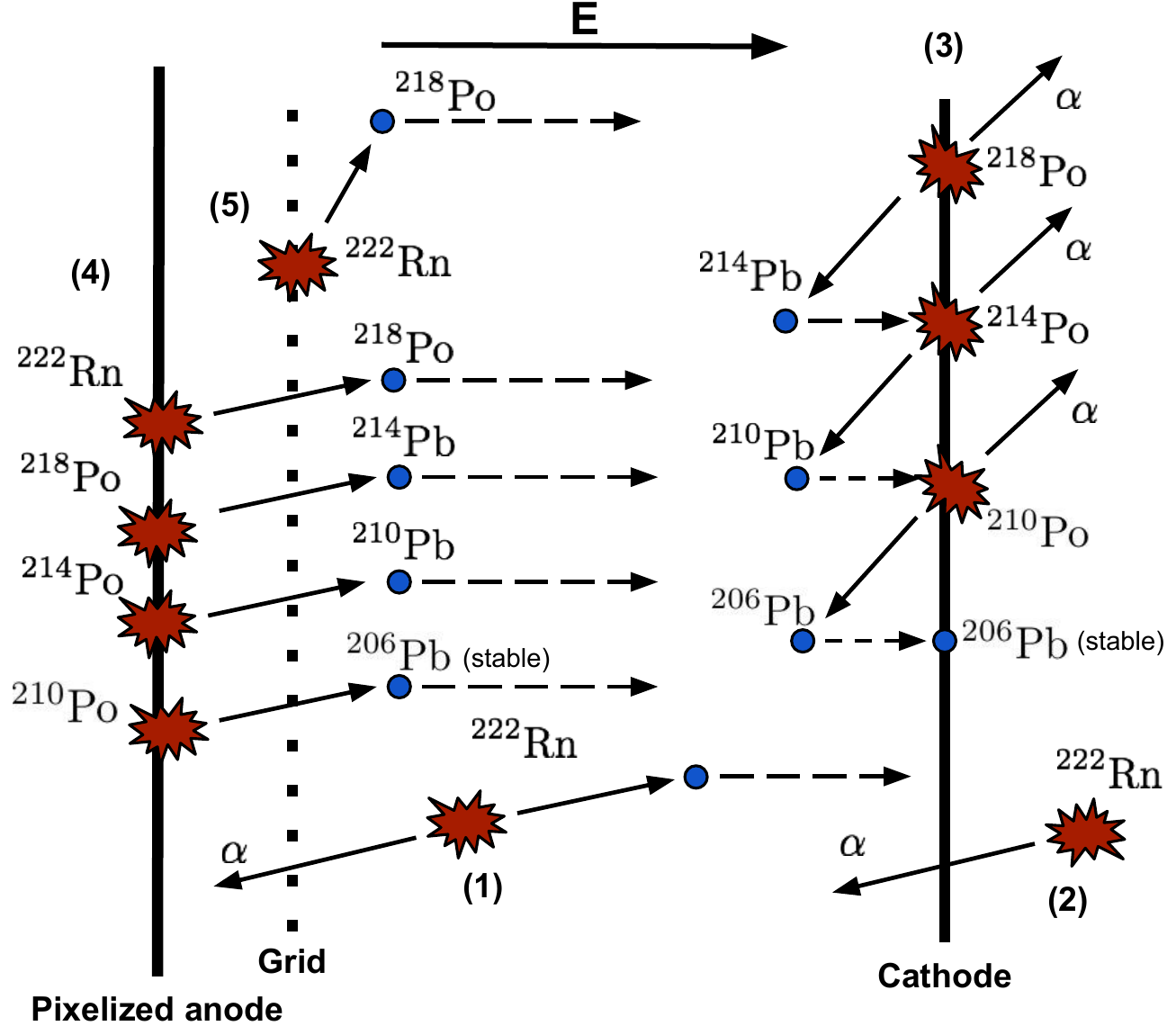}
		\caption{ This schematic diagram illustrates the daughter RPR event spatial distribution. The $\alpha$-decays are represented by the red polygons. Daughter nuclei and their paths are respectively represented by blue dots and plain arrows. Dashed arrows represent the daughter nucleus migration. See section~\ref{sec-SourceId} for more details.}
		\label{fig:origin}
\end{figure}
	
	\item[2) Through cathode events ] 
	A SRIM~\cite{ZieglerJ.F.;Biersack1985} simulation shows that \ag-particles with kinetic energies ranging from 5.5 to 8.8~MeV (see table \ref{tab:Progeny}) can pass through a 12~\textmu m thickness mylar cathode and reach the other chamber with kinetic energies going up to hundreds of~keV. For such events, we observe coincidences of signals from the two chambers.
		
	\item[3) Cathode events] As mentioned before in the case of RPR volume events, each daughter nucleus produced in the drift electric field is collected on the cathode surface.
	Moreover, radon isotopes from the detector gas can also be fixed on the cathode surface by adsorption.
	At the cathode surface there is an accumulation of radon progeny events.
	While an \ag-decay occurs at the cathode surface, there are two cases to consider: i)~if the \ag-particle is emitted in the direction of the gas volume, as mentioned before, the ionization energy deposition saturates the preamplifier,
	ii)~if the $\alpha$-particle is absorbed in the matter, only the recoil of the daughter nucleus is detected. The recoils of the daughter nuclei from \Nucl{Rn}{222} and \Nucl{Rn}{220} progeny have kinetic energies from 100 to 170~keV, as shown in the table~\ref{tab:Progeny}, with simulated track lengths ranging from 650 to 900~\textmu m.
	In addition, there is an important difference between the measurable ionization energy $E_{recoil}^{ioni}$ and the kinetic energy $E_{recoil}^{kin}$ for low energy NRs defined by the Ionization Quenching Factor ({\it IQF}).
The  {\it IQF} of a NR is defined by the ratio of the measured ionization energy $E_{recoil}^{ioni}$ and its kinetic energy $E_{recoil}^{kin}$ or the ionization energy released by an electron of the same kinetic energy. 
This factor decreases rapidly with decreasing kinetic energies of NRs. It depends on the mass of the NR and on the gas and pressure, as shown in~\cite{Guillaudin2012}.
In our case, a SRIM simulation gives an {\it IQF} of about 40\% for a heavy nucleus such as \Nucl{Po}{218} at 100 keV. Taking into account this correction from SRIM, the RPR events should release an ionization energy from 38 to 58~keVee as shown in the table~\ref{tab:Progeny}.
	In this case, the daughter NR is associated with an \ag-particle (plain arrow) passing through the thin mylar cathode and reaching the other chamber. These events will be in coincidence for the data analysis.
	When a radon progeny nucleus recoils in the active volume, it is collected again at the cathode surface by the drift electric field. Then, at the cathode surface, all the \Nucl{Rn}{222} and \Nucl{Rn}{220} progeny elements are collected. 
	
	\item[4) Anode events] The Micromegas PCB and strips contain the most important \Nucl{U}{238} and \Nucl{Th}{232} pollution. 
	Consequently they are the major sources of radon internal emanations.
	In the 256~\textmu m amplification space, the multiplication of ionization electrons depends on their positions and it affects the energy measurement of an event passing 
	through the amplification space. The impact of the gain variation on the energy measurement is discussed in section~\ref{sec:AnodeAndCathodeEvents}.
	At the anode level we expect two types of \ag-particle and/or RPR events: bulk events from the decay of \Nucl{U}{238} and \Nucl{Th}{232} and surface events from the radon isotopes emanations.
	In the case of bulk events, the NR leaves the surface with a reduced energy and reaches the amplification space. In some cases, it is possible for a bulk event to pass through the mesh and reach the drift space. 
	In the case of surface events, the daughter NRs pass through the mesh reaching the drift space.
	In any of these cases, the ionization energy measurement misestimates the total ionization energy deposition due to the gain variation through the amplification space.
	The daughter NR from the \ag-decay is collected either at the mesh or at the cathode surface.

	\item[5) Mesh events] As mentioned before, radon isotopes can be fixed on the Micromegas mesh wire surfaces by adsorption. The mesh is a woven stainless steel thread of 18 \textmu m diameter wires. While \ag-decays occur at wire surfaces and daughter recoils enter on the active volume, the associated \ag-particles have a probability of 10\% to pass throughout the wire. The energy released by these \ag-particles in the amplification zone will be in that case added to daughter nuclei deposition. The daughter NR is collected at the cathode surface.
				
\end{description}

All RPR events in the detector contribute to the accumulation of RPR at the cathode surface. 
Cathode and passing through the cathode events can be identified using the coincidence between both chambers.
The main feature of the anode and mesh events is the misestimation of the ionization energy either due to gain variation in the amplification space or due to the addition of an associated \ag-particle energy fraction.

\section{ Radon progeny Recoils evidences}
\label{seq:RPR}

\subsection{\Nucl{Rn}{222} pollution evidence}
\label{sec-pollutionEv}

\begin{figure}[tbp]
\centering
		\includegraphics[width=0.8\linewidth]{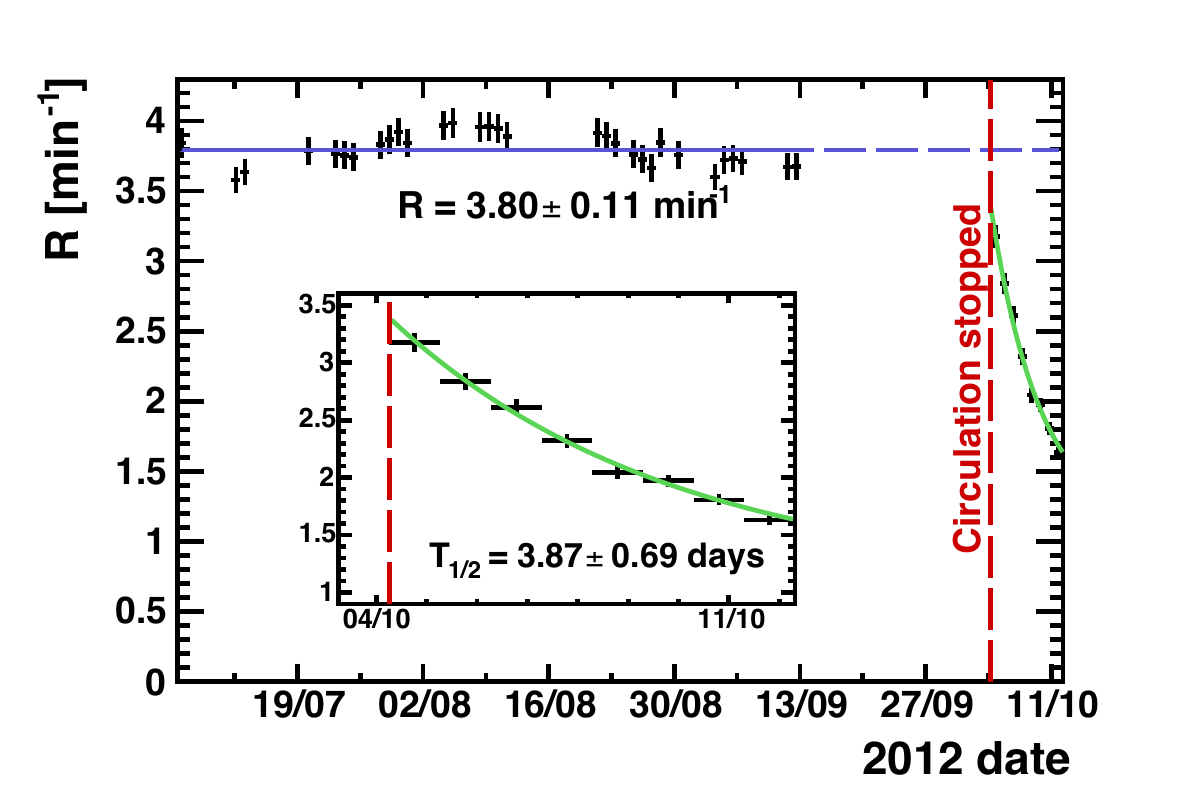}
		\caption{ The saturation event rate  covering the first data run and a zoom of the last twelve days. The circulation system was stopped on October~3$^{rd}$, 2012, as represented by the orange dashed line. The green line represents the exponential fit of the event rate. The measured half-life is $3.87\pm0.69\,\mathrm{days}$, it is compatible with the \Nucl{Rn}{222} half-life (3.8 days).}
	\label{fig:alpha}
\end{figure}

The \ag-particle rate was monitored selecting the 3D tracks saturating the preamplifier in the 2012 data set. This selection includes \ag-particles from the active volume and from the other chamber through the mylar cathode.
Figure~\ref{fig:alpha} shows the 2012 \ag-particles rate. From July $13^{\mbox{th}}$ to September $14^{\mbox{th}}$, the \ag-rate was rather constant at $3.80 \pm 0.11\,\mathrm{min^{-1}}$. 

On October 3rd, the gas circulation was switched off (red dashed line). We observed an exponential reduction of the event rate. We model it by the sum of a constant distribution $c$ and a decreasing exponential with a $T_{1/2}$ half-life:
\begin{equation}
R(t) = R_{0} \exp{\left(- \frac{t}{\tau}\right)}+c \mbox{, where } \tau = T_{1/2}/\ln(2)
\end{equation}
It indicates an external pollution of the gas mixture from the circulation gas system and an intrinsic pollution from the materials.
By fitting the event rate (green line), we obtained a $T_{1/2} = 3.87\pm 0.69\,\mathrm{days}$    Half life which is compatible with the 3.8 days half-life of the \Nucl{Rn}{222}. The constant $c=1.1\pm 0.4\,\mathrm{min^{-1}}$ estimates the intrinsic saturation event rate from the materials pollution.
We identified the external source of contamination in the gas circulation system as a small leak in the circulation pump. During this data taking, we measured a $1.7\times 10^{-4}\,\mathrm{mbar.L/s}$ global leak rate. This leak injected \Nucl{Rn}{222} and \Nucl{Rn}{220} from the air of the cavern into the circulation loop. According to calibration data, no gain degradation was observed thanks to the presence of oxygen and humidity filters. After October~3$^{\mathrm{rd}}$,  the circulation pump leak was fixed and the global leak rate was reduced by a factor of 4.5. After this operation, we measured a saturation event rate of $0.9\pm 0.3\,\mathrm{min^{-1}}$ which is compatible with the estimated intrinsic saturation event rate showing that our background is now dominated by  internal sources of contamination of the detector. In addition, the impact of this leakage is as well visible on the total event rate as shown by the table\ref{tab-summRun}.

The exponential reduction of the event rate was dominated by the contribution from the \Nucl{Rn}{222} half-life. From this measurement, we could not conclude about the contribution from the \Nucl{Rn}{220} half-life. In general, the \Nucl{Rn}{220} contribution to the RPR event progeny is neglected~\cite{Beringer2012}. Due to its shorter period (55~s vs 3.8~days) the impact of emanating \Nucl{Rn}{220} from materials is much smaller than \Nucl{Rn}{222} contributions.  This hypothesis was supported by a dedicated one-month measurement performed with 700~mbar of pure $\mathrm{CF_4}$. The \ag-particle spectroscopy showed no contribution coming from the $8.8\,\mathrm{MeV}$ alpha particle from the \Nucl{Po}{212} decay in the $\alpha$-particle spectrum of the intrinsic background.

\begin{figure}[tbp]

\centering
		\includegraphics[width=\linewidth]{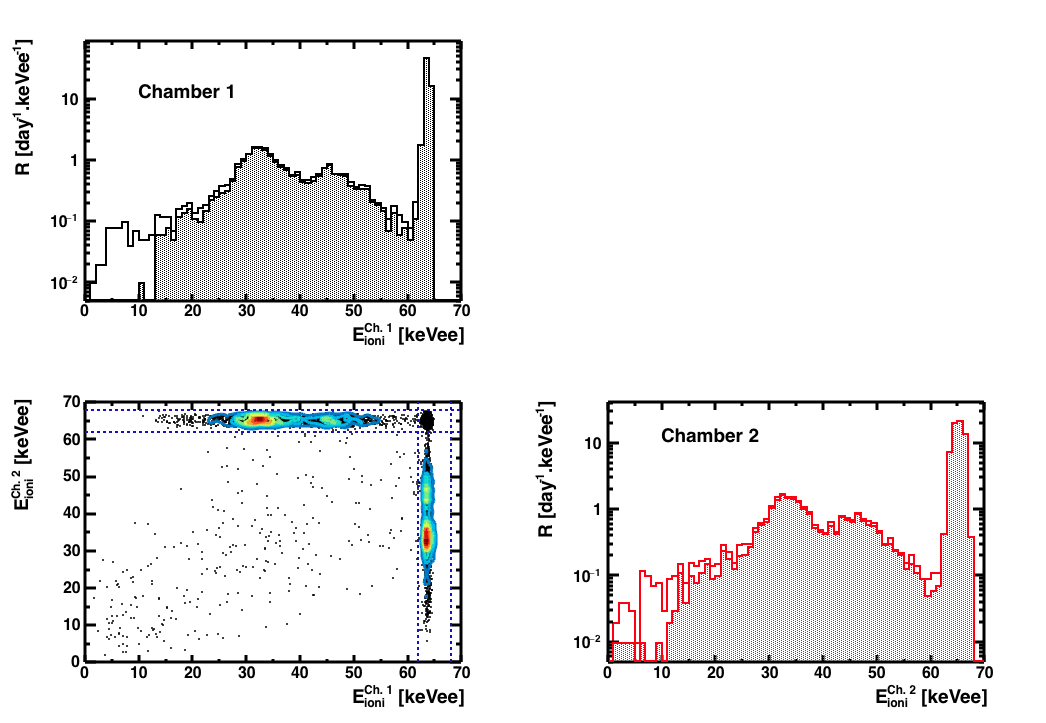}
		\caption{ 2D distribution with contour levels of the energy seen in each chamber for in-coincidence events (bottom left panel) and 1D marginalization (top left - bottom right). The regions delimited by the blue dashed lines represent events associated with saturation in the other chamber. 
		The black (red) spectrum represents the chamber~1 (2) measured energies and the filled area corresponds to events associated with saturation as mentioned in section~\ref{sec:Signatures}. 
		}
		\label{fig:Coinc}
\end{figure}

\subsection{In coincidence events}
\label{sec:Incoinc}

The chamber clocks were synchronized with a 40~ns precision, allowing chamber coincidence searches. 
As explained in section~\ref{seq-Origin}, we are expecting two different types of events in coincidence: cathode events and passing through cathode events.
We consider two events in coincidence if there are separated by less than $11.6$ \textmu s. It corresponds to the time required to travel the distance between  the cathode and the anode (25 cm) for primary ionization electrons. Indeed, in the case of a cathode event as described before (case 3), the \ag-particle can pass through the gas volume releasing part of its energy close to the anode while the daughter NR releases its energy close to the cathode. In that case, we measure a prompt signal from the \ag-particle and a delayed signal from the daughter nucleus with a $11.6$ \textmu s delay.

Considering that the \Nucl{Rn}{220} contribution is negligible with respect to the \Nucl{Rn}{222} contribution, we expect to find four peaks on the "in coincidence" event energy spectrum from cathode events: \Nucl{Po}{218}, \Nucl{Pb}{214}, \Nucl{Pb}{210} and \Nucl{Pb}{206} NRs.
In a first approximation, we can also neglect the \Nucl{Pb}{206} contribution  due to the long half-life of \Nucl{Pb}{210}.

At the cathode surface, the \Nucl{Po}{218} contribution comes from the attachment of \Nucl{Rn}{222} while the other contributions come from the collection of the \ag-decay daughter nuclei due to the drift electric field. In this context, the \Nucl{Po}{218} population at the cathode surface comes from the collection of \Nucl{Po}{218} after \ag-decays.
If we neglect the \Nucl{Rn}{222} attachment, two main contributions remain: \Nucl{Pb}{214} and \Nucl{Pb}{210}.

\begin{figure}[tbp]
\centering
		\includegraphics[angle=90,width=0.7\linewidth]{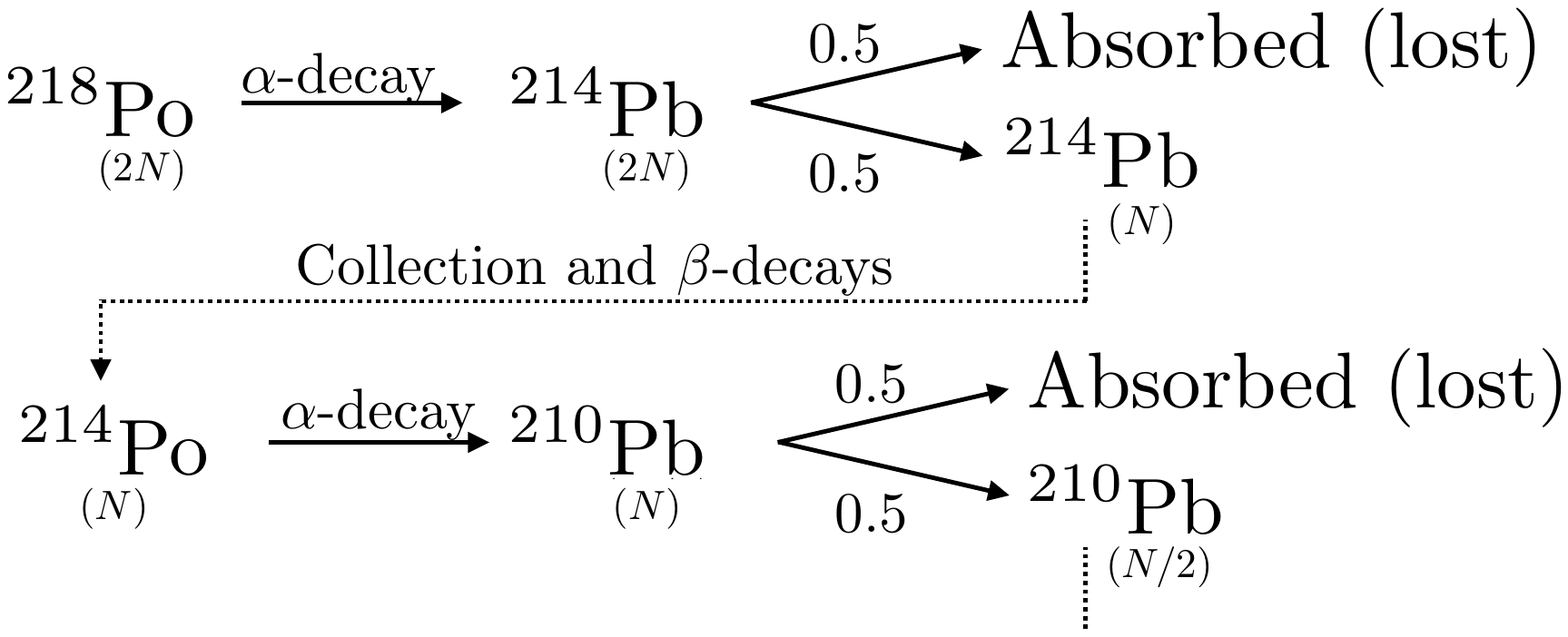}
		\caption{Illustration of the relative contribution of \Nucl{Pb}{214} and \Nucl{Pb}{210} and the evolution of their population at the cathode surface. This figure illustrates why the \Nucl{Pb}{210} contribution amplitude is 2 times smaller than the \Nucl{Pb}{214} one.}
		\label{fig:Population}
\end{figure}

Figure~\ref{fig:Population} illustrates the relative contribution of \Nucl{Pb}{214} and \Nucl{Pb}{210} and the evolution of their population at the cathode surface. 
Considering $2N$ \Nucl{Po}{218} nuclei at the cathode surface. After \Nucl{Po}{218} \ag-decay, \Nucl{Pb}{214} daughter nucleus has a 0.5 probability to be absorbed by the surface. The number of \Nucl{Pb}{214} emitted in the gas is $N$. Assuming a 100\% efficiency for the \Nucl{Pb}{214} collection, the number of \Nucl{Po}{214} at the cathode surface is $N$ after two $\beta$-decays. As previously, after \Nucl{Po}{214} \ag-decay, \Nucl{Pb}{210} daughter nucleus has a 0.5 probability to be absorbed by the surface. Thus, the number of \Nucl{Pb}{210} emitted in the gas is $N/2$. In conclusion, the \Nucl{Pb}{210} contribution amplitude is 2 times smaller than the \Nucl{Pb}{214} one due to the population reduction by the surface absorption.

In conclusion, we can expect two well-defined peaks from RPR cathode events, one flat distribution from the \ag-particles passing through the cathode and an important saturation from \ag-particles which release more than 62~keVee.

The panels on the top left - bottom right diagonal of figure~\ref{fig:Coinc} present the "in coincidence" event energy spectra in both chambers measured in 2013. 
These spectra show two peaks at roughly 33 and 45~keVee and saturation over 62~keVee. 

The left bottom panel presents the 2D distribution of the energy seen in each chamber for "in coincidence" events. 
We can clearly identify four regions delimited by the blue dashed lines:
\begin{itemize}[leftmargin=0.4cm]

\item The "double saturation" region corresponds to events with $E^{\mathrm{Ch. 1}}_{\mathrm{ioni}}$ and $E^{\mathrm{Ch. 2}}_{\mathrm{ioni}}>62\,\mathrm{keVee}$. It represents of 44\% of the total "in coincidence" event sample. The events belonging to this region are \ag-particles passing through the 12 \textmu m mylar cathode and saturating the preamplifiers in both chambers.

\item There are two "one saturation" regions ($E^{\mathrm{Ch. 1}}_{\mathrm{ioni}}$ or $E^{\mathrm{Ch. 2}}_{\mathrm{ioni}}>62\,\mathrm{keVee}$).   These events (52\% of the total "in coincidence" event sample) correspond to RPR events associated with an \ag-particle detected in the other chamber or not fully detected \ag-particles in both chambers (in/out-going \ag-particles).
In these regions, we can clearly identify the contour levels of the two peaks shown by the marginalized energy distributions at 33 and 45~keVee. 

\item The non-saturation region corresponds to events with $E^{\mathrm{Ch. 1}}_{\mathrm{ioni}}$ and $E^{\mathrm{Ch. 2}}_{\mathrm{ioni}}<62\,\mathrm{keVee}$. This event sample constitutes 4\% of the total. These events correspond to RPR events with a non-saturating \ag-particle such as an in/out-going \ag-particles, or a not fully detected RPR event.

\end{itemize}

This figure shows that most of "in coincidence" events are associated with an energy saturation: {\it i.e.} with an \ag-particle. It is illustrated by filled areas which represent the events associated with a saturated event. More than 95\% of the "in coincidence" events are in the saturation regions.

\begin{figure}[tbp]
\centering
		\includegraphics[width=0.8\linewidth]{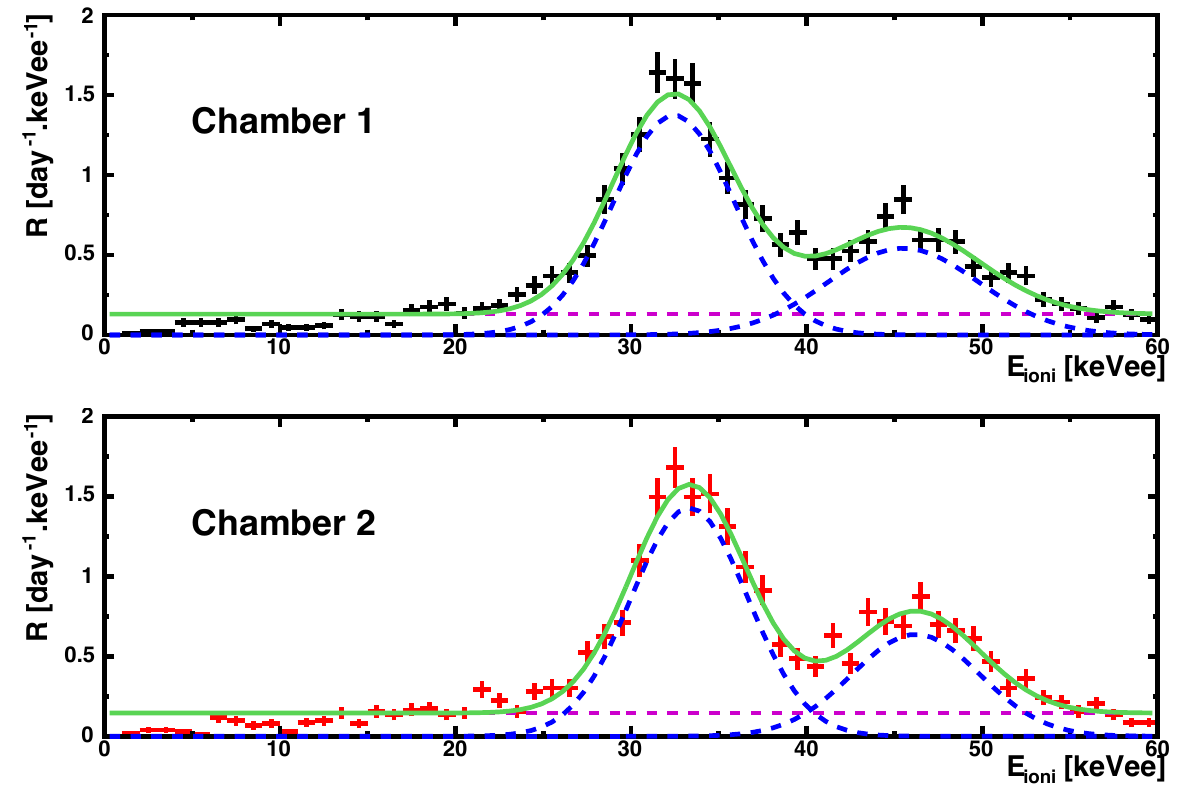}
		\caption{Zoom from 0 to 60~keVee of the "in coincidence" event energy spectra in the chamber~1 (top panel) and 2 (bottom panel). The plain green line corresponds to the fit result by the sum of two gaussians and a constant (see equation~\ref{eq:fit}). The dashed blue and Magenta lines represent the individual contributions. }
		\label{fig:Coinc_Fit}
\end{figure}

Figure~\ref{fig:Coinc_Fit} shows a zoom of these energy spectra from the threshold in ionization energy to 62~keVee.
These energy spectra show only two main gaussian contributions as expected before: \Nucl{Pb}{214} and \Nucl{Pb}{210}. 
This observation supports our hypothesis about the \Nucl{Rn}{222} adsorption contribution.

\begin{table}
\centering
	\begin{tabular}{|c|c|c|c|}
	\hline
	\rule[-0.2cm]{0cm}{0.55cm} Parameter &Unit & chamber~1 & chamber~2 \\
	\hline
	\hline
	$ Cst $		& [$\mathrm{day^{-1}.keVee^{-1}}$]	& $0.13 \pm 0.01$  	& $0.14 \pm 0.01$ \\
	\hline
	$A_1$ 		& [$\mathrm{day^{-1}.keVee^{-1}}$]	& $11.8 \pm 0.4$ 	& $ 11.9 \pm 0.5$ \\
	$\mu_1$ 		& [keVee]						& $ 32.4 \pm 0.2$ 	& $33.3 \pm 0.1$\\
	$\sigma_1$	& [keVee]						& $ 3.4 \pm 0.2 $ 	& $ 3.3 \pm 0.1$ \\
	\hline
	$A_2$ 		& [$\mathrm{day^{-1}.keVee^{-1}}$]	& 5.8$ \pm 0.4$ 	& $ 5.8\pm 0.4$ \\
	$\mu_2$ 		& [keVee]						& $ 45.5 \pm 0.4$ 		& $46.2 \pm 0.3$ \\
	$\sigma_2$ 	& [keVee]						& $ 4.3 \pm 0.3$  		& $ 3.6\pm 0.2$ \\
	
	\hline	
	\end{tabular}	
	\caption{Parameters of the "in coincidence" energy spectra fits in figure~\ref{fig:Coinc_Fit} by the equation~\ref{eq:fit}.  }
	\label{tab:FitInCoinc}
\end{table}

The energy spectra were fitted using the sum of a constant and two gaussians from 15 to 60~keVee:
\begin{equation}
f(E_{\mathrm{ioni}}) = Cst + \sum_{i=1}^{2}{\frac{A_i}{\sigma_i\sqrt{2\pi}}\exp{\left(  -\frac{\left( E_{ioni}-\mu_i  \right)^2}{2\sigma_i^2}\right)}}.
\label{eq:fit}
\end{equation}
In figure~\ref{fig:Coinc_Fit}, the green line represents the fit result, blue and magenta dashed lines the individual contributions. Table~\ref{tab:FitInCoinc} presents the values of the fit parameters for both chambers.
These two peaks, measured at $32.85\pm0.28$ and $45.85\pm0.5\,\mathrm{keVee}$, taking into account the mean values between the two chambers, correspond respectively to the ionization energy released by \Nucl{Pb}{214} and \Nucl{Pb}{210} NRs. The measured {\it IQF} values are $Q(\mathrm{^{214}Pb}) = 29.2\pm 0.1\%$ for \Nucl{Pb}{214} and $Q(\mathrm{^{210}Pb}) = 31.3 \pm 0.2\%$ for \Nucl{Pb}{210} showing respectively a $25 \%$ and 22\% SRIM {\it IQF} overestimation (see table~\ref{tab:Progeny} for simulated values). 
This is consistent with previous results published by our team on {\it IQF} measurements on several gas mixtures and for several ions~\cite{Guillaudin2012}.

\begin{figure}[tbp]
\centering
		\includegraphics[width=0.8\linewidth]{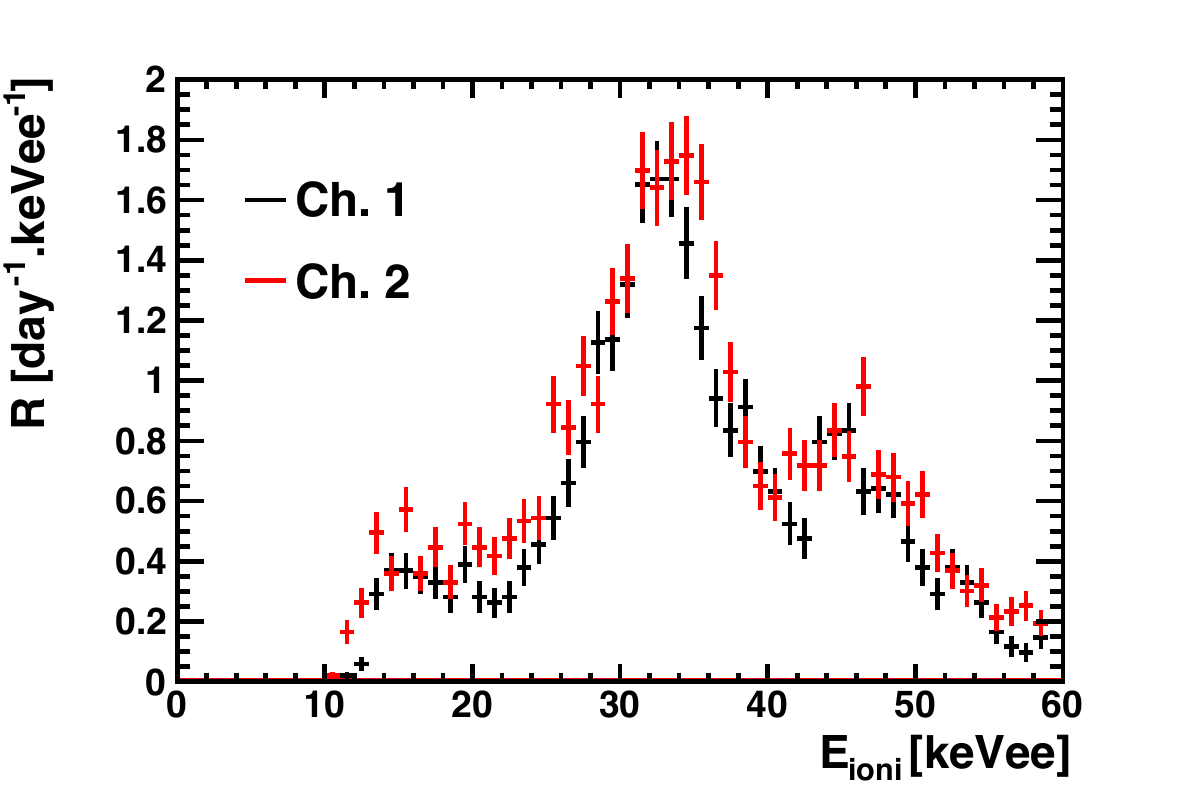}
		\caption{ Energy spectra measured in 2013 by the chamber~1 (black line) and the chamber~2 (red line). It should be pointed out that each event represented on these spectra has its own 3D track associated. These distributions were obtained applying the low energy ER/NR discrimination.}
		\label{fig:DiscriApplication}
\end{figure}

The ratios of the two peak amplitude are $2.02 \pm 0.12$ for the chamber~1 and $2.04 +/- 0.16$ for the chamber~2  supporting our statement described above about the  \Nucl{Po}{218} and \Nucl{Pb}{206} contributions. 

\section{RPR recoil selection}
\label{sec-RPR_recoil_selection}

In order to reject the ER background from dark matter searches data, we developed an original ER/NR discrimination method described in~\cite{Riffard2016mgw}. We placed a MIMAC mono-chamber detector on a monochromatic neutron field allowing us to acquire two specific data sets: (ER and NR) and ER only. We applied a Boosted Decision Tree (BDT) algorithm on these two data sets. It gives a $10^5$ ER rejection power. As discussed in~\cite{Riffard2016mgw}, by using this method the detector efficiency is not directly accessible. Then we have developed a Monte-Carlo simulation of the MIMAC readout that is able to reproduce our observables. The application of the BDT analysis on the Monte Carlo shows a $86.49 \pm0.17$\% NR efficiency considering the full energy range and $94.67 \pm 0.19$\% considering a 5 keV lower threshold with an $10^5$ ER rejection power.
This method was applied on the 2013 data run (103~days) and figure~\ref{fig:DiscriApplication} presents the resulting NR energy spectra. 
After the application of this analysis, we can consider the ER event contamination in our data as negligible.
These energy spectra show two peaks at 33 and 46~keVee, in both chambers (red line chamber~1 and black line chamber~2) as already observed in the "in coincidence" energy spectra. 
However, the observed shapes are different from the "in coincidence" spectrum shapes, especially below 25~keVee. 
These differences are due to the fact that these spectra contain all contributions from RPR events described in section~\ref{sec:Signatures}.
We measure a total RPR event rate of $29.3 \pm 0.5\,\mathrm{day^{-1}}$ in chamber~1 and $34.8 \pm 0.6\,\mathrm{day^{-1}}$ in chamber~2. 

The fact that some progeny events are passing our BDT cuts shows that their reduction is needed for dark matter searches. The reduction of this background can be done by:
\begin{itemize}
\item A screening of the materials in order to reduce their radioactivity and radon emanations.
\item A rejection of the RPR by the coincidence between the different chambers. This method is limited by the thickness of the cathode that can prevent the coincidence.
\item A fiducialisation on z axis of the detector. As these events are located at the cathode and anode levels, a cut on the z-coordinate of the events must allow to suppress their contribution. A new signal, from the cathode, produced by the primary electrons drift will be added~\cite{couturier_2017_cathode}.
\end{itemize}
In section~\ref{secMPD}, we present a new observable exploiting the electron diffusion to identify the RPR events position along the z-coordinate.

\begin{figure}[tbp]
\centering
		\includegraphics[width=0.8\linewidth]{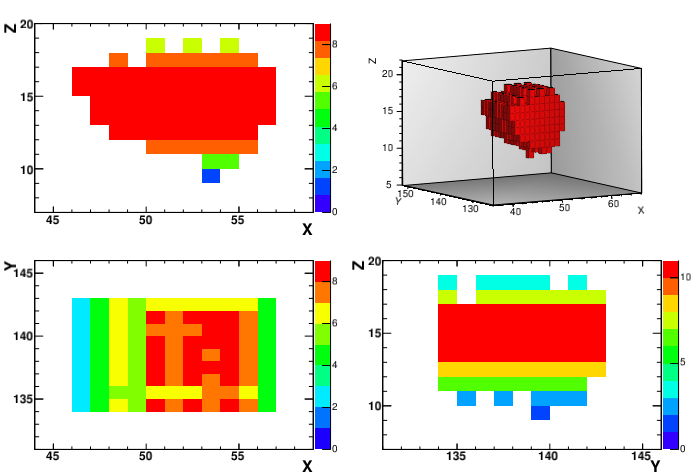}\\
		\caption{ Projections of a 41.1 measured NR track in the (X,Z), (Y,Z) and (X,Y) planes and 3D reconstructions. The Z axis is in units of time slice (20 ns) and the X and Y axis in strip number (424\textmu m pitch). The color scale corresponds to the number of strips fired on the time slice.}
		\label{fig:TrackCathode}
\end{figure}
\begin{figure}[tbp]
\centering
		\includegraphics[width=0.8\linewidth]{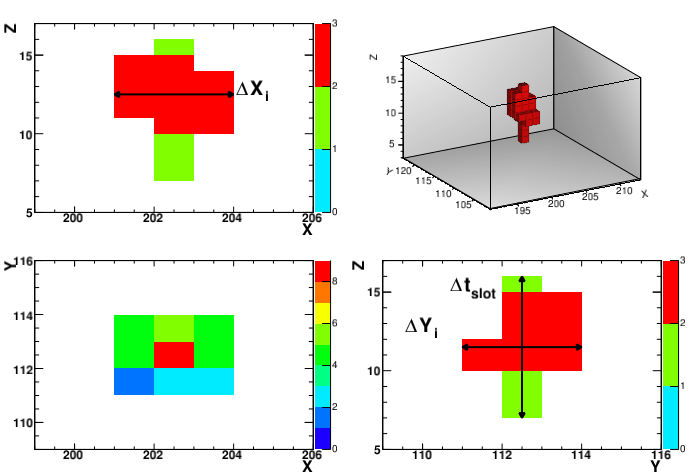}
		\caption{ Projections of a 36.4~keVee measured NR track in the (X,Z), (Y,Z) and (X,Y) planes and 3D reconstructions. The Z axis is in units of time slice (20 ns) and the X and Y axis in strip number (424\textmu m pitch). The color scale corresponds to the number of strips fired on the time slice. The horizontal arrows represent the width along the X/Y axis of the time slice $i$: $\Delta X_i$/$\Delta Y_i$ . The vertical arrow represents the slot duration $\Delta t_{slot}$.}
		\label{fig:TrackAnode}
\end{figure}

\section{Radon progeny recoil position identification}
\label{sec-SourceId}


\subsection{The Mean Projected Diffusion observable}
\label{secMPD}

As seen in section~\ref{seq:RPR}, the RPR events occur at different positions in the detector. The position determination in the ($X,Y$) plane is made using the pixelated Micromegas readout, while the $z$ coordinate identification needs to use the information given by the electron diffusion. 
Indeed, the electron diffusion in the drift space is directly related to the $z_0$ coordinate via the probability density function of charge distribution on the anode plan. 
The transverse/longitudinal standard deviation of the charge dispersion $\sigma_{T/L}$ follows a square root dependency on the distance of the track to the anode ($z_0$):  $\sigma_{T/L}= D_{T/L}\sqrt{z_0}$, where $D_{T/L}$ is the diffusion coefficient. In order to obtain the distance of the track to the anode, we take the distance with respect to the center of the track.
The diffusion coefficients $D_{T/L}$ at such pressure and electric field have been calculated using Magboltz~\cite{Biagi1999}, giving the following values:
\begin{equation*}
	\left\{
	\begin{array}{c}
		D_T = 237.9 \,\mbox{\textmu m}/\mathrm{\sqrt{cm}}\\
		D_L = 271.5 \,\mbox{\textmu m}/\mathrm{\sqrt{cm}}\\
	\end{array}
	\right.
\end{equation*}

Figures~\ref{fig:TrackCathode}~and~\ref{fig:TrackAnode} show respectively the NR tracks of a 41.1~keVee cathode event and a 36.4~keVee anode event. These figures illustrate the  impact of the ionization  electron diffusion on the track topology. 
It affects the measurement of the track length and the track width ($\Delta X/\Delta Y$) depending on the $z_0$ track coordinate. As illustrated in figure~\ref{fig:TrackAnode}, close to the anode, the electron diffusion is negligible.
However, as illustrated by figure~\ref{fig:TrackCathode}, close to the cathode the electron diffusion is maximal. In that case, the edges of the primary electron density from cathode events are not completely sampled by the pixelated anode due to the strip thresholds. It implies that the measured track length after a diffusion deconvolution is misestimated.

\begin{figure}[tbp]
\centering
		\includegraphics[width=0.8\linewidth]{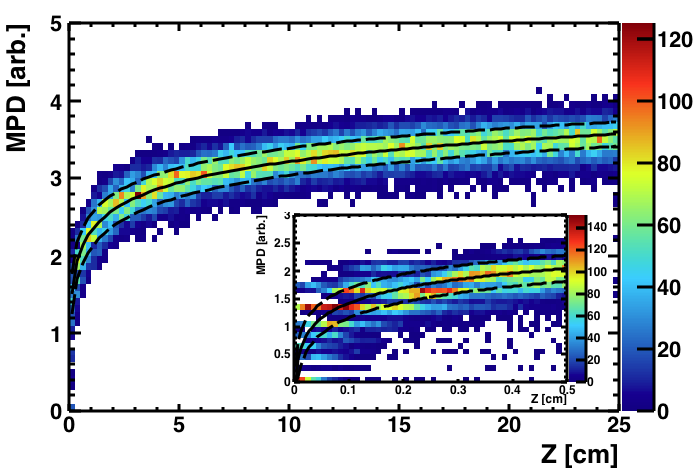}
		\caption{Simulated MPD distribution as a function of the track $z$ coordinate. This simulation shows 112.3~keV \Nucl{Pb}{214} tracks generated by SRIM~\cite{ZieglerJ.F.;Biersack1985} and coupled to the MIMAC detector simulation. The black plain and dashed lines represent respectively the mean and the standard deviation of the MPD as a function of the ionization energy. }
	\label{fig:PoMPD}
\end{figure}

The 2D projections on the (X,Z) and (Y,Z) planes of the bottom panel  show the definition of $\Delta X_i/\Delta Y_i$, the width of the $i^{st}$ time slice along the $x/y$ axis. 
Then, in order to estimate the transverse contribution of the diffusion, we defined an observable called Mean Projected Diffusion \footnote{MPD} as:
\begin{equation}
MPD = \log\left({\overline{\Delta X}\times \overline{\Delta Y}}\right),
\end{equation}
where $\overline{\Delta X}$ and $\overline{\Delta Y}$ are respectively the mean value of $\Delta X_i$ and $\Delta Y_i$ $( i= 1, N)$. As an example, the events show by figures~\ref{fig:TrackCathode}~and~\ref{fig:TrackAnode} have respectively a 4.03 and a 1.21 MPD values.

Figure~\ref{fig:PoMPD} presents a simulated distribution of 112.3~keV \Nucl{Pb}{214} events in the $(z,\mathrm{MPD})$ plane. 
This simulation is based on \Nucl{Pb}{214} tracks generated by SRIM with a kinetic energy of 112.3~keV on MIMAC gas mixture at 50 mbar. It included the ionization electrons diffusion using a diffusion coefficient estimation from Magboltz as mentioned before.
We can see that MPD increases as the square root of the track $z$ coordinate. The diffusion variation along the track is negligible for small tracks ($ \sim 0.5 - 10$~mm). As track length and mean width depends on the NR energy, the MPD depends on energy too. 
In conclusion, using this observable, we can identify the track position of the events in the detector. This observable allows one to discriminate anode events from volume or cathode events.
For \Nucl{Pb}{214} NRs of 112.3~keV, cathode and volume events have a MPD value higher than 2.3, while anode events have a value lower than 2.3 for the lengths of NRs being smaller than 1 mm, see insert in figure~\ref{fig:PoMPD}.

\subsection{Anode and cathode radon progeny recoils}
\label{sec:AnodeAndCathodeEvents}
\begin{figure}[tbp]
\centering
		\includegraphics[width=0.8\linewidth]{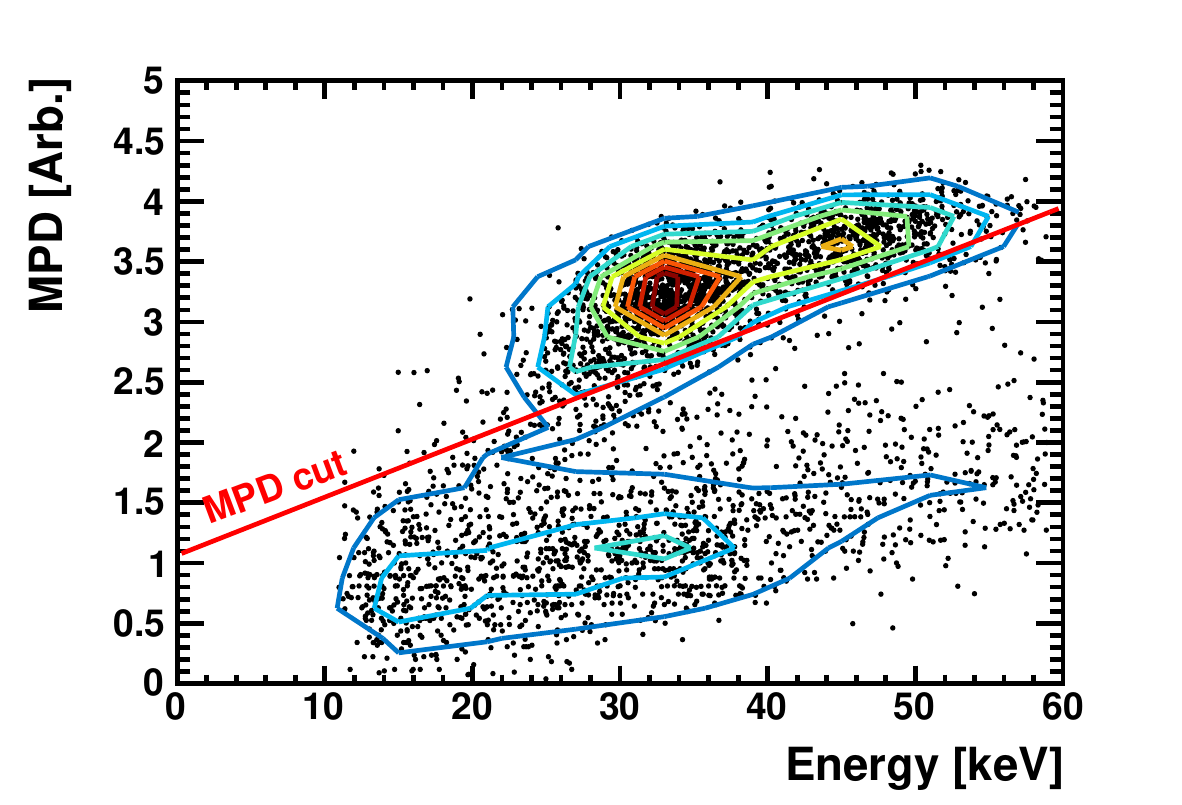}
		\caption{ MPD distribution as a function of the event ionization energy in the chamber~2. These distributions were obtained applying the ER/NR discrimination. The red line corresponds to the cut on the MPD.}
		\label{fig:MPD}
\end{figure}

\begin{figure}[tbp]

\centering
		\includegraphics[width=\linewidth]{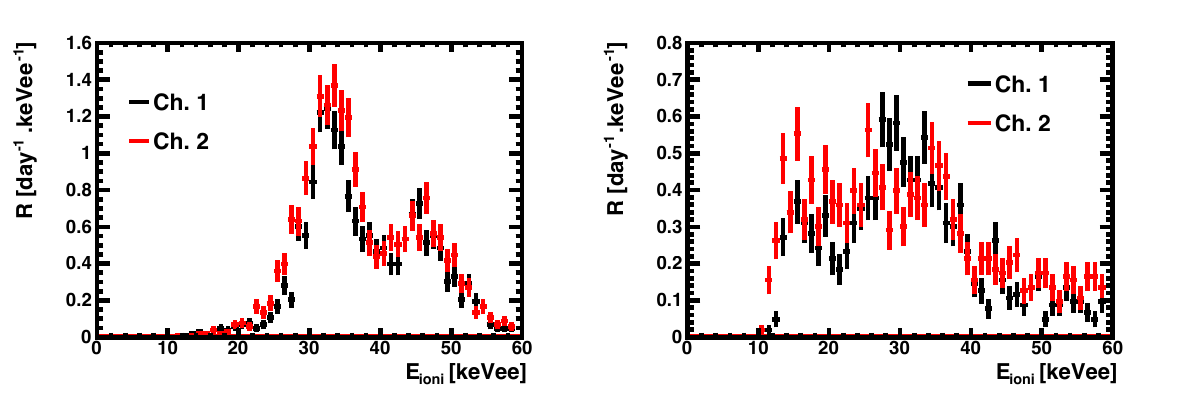}
		\caption{The energy spectra measured in 2013 by the chamber~1 (black line) and 2 (red line) after ER/NR discrimination with a cut on the event MPD value. The left panel represents the energy spectra of volume event and cathode events (above the MPD cut), and the right panel the energy spectra of anode and mesh events (below the MPD cut). }
		\label{fig:DiscriApplicationMPD}
\end{figure}

Figure~\ref{fig:MPD} shows the measured MPD distribution as a function of the ionization energy $E_{ioni}$ in the chamber~2. 
In this figure, two regions can clearly be identified. Using the in-coincidence data, we performed a cut on MPD keeping 99\% of the in-coincidence events in the upper region represented by the red line in figure~\ref{fig:MPD}. This cut was optimized in order to take into account the energy dependence of the MPD.
The upper and lower regions correspond respectively to volume and cathode events and to anode and mesh events.
Figure~\ref{fig:DiscriApplicationMPD} shows the energy spectra of the two chambers with an MPD above the cut (left panel) and with an MPD below the cut (right panel).

In the region above the MPD cut the energy spectra show two peaks observed at 33 and 46~keVee. It confirms that these events are related to RPR events from the cathode, shown previously on  coincidence spectra, with a measured event rate of $17.3 \pm 0.4\,\mathrm{day^{-1}}$ in chamber~1 and $20.9 \pm 0.4\,\mathrm{day^{-1}}$ in chamber~2. These energy spectra were fitted from 25 to 60~keVee using the same model as in section~\ref{sec:Incoinc}. In table~\ref{tab:FitSpectreMPD} the fit parameter values are given.

The fitted peak positions match with the peak positions previously fitted in section~\ref{sec:Incoinc}. It confirms that the observed events are related to RPR events from the \Nucl{Rn}{222} decay chain.
The ratios of the two peak amplitudes are $2.18 \pm 0.23$ for the chamber~1 and $2.01 \pm 0.16$ for the chamber~2. These values are still compatible with a ratio of 2 supporting our statement.

The right panel of figure~\ref{fig:DiscriApplicationMPD} shows the ionization energy distribution of events below the MPD cut, events from the mesh and anode. In these spectra, no clear peak structures are observable, this is due to an incomplete  electron avalanche of events crossing the amplification space. This anode and mesh backgrounds contribute to the global background with a measured event rate of $12.0 \pm 0.3\,\mathrm{day^{-1}}$ in chamber~1 and $14.0 \pm 0.4\,\mathrm{day^{-1}}$ in chamber~2.

\begin{table}
\centering
	\begin{tabular}{|c|c|c|c|}
	\hline
	\rule[-0.2cm]{0cm}{0.55cm} Parameter &Unit & chamber~1 & chamber~2 \\
	\hline
	\hline
	$ Cst $		& [$\mathrm{day^{-1}.keVee^{-1}}$]	& $10.1 \pm 1$  	                                  & $1.1\pm 0.8$ \\
	\hline
	$A_1$ 		& [$\mathrm{day^{-1}.keVee^{-1}}$]	& $1.02\times10^3 \pm 0.05\times10^3$ 	& $ 1.21\times10^3 \pm 0.5\times10^3$ \\
	$\mu_1$ 		& [keVee]						& $ 32.5 \pm 0.2$ 	                                  & $32.9 \pm 0.2$\\
	$\sigma_1$	& [keVee]						& $ 3.3 \pm 0.2 $ 	                                  & $ 3.9 \pm 0.2$ \\
	\hline
	$A_2$ 		& [$\mathrm{day^{-1}.keVee^{-1}}$]	& $4.7\times10^2  \pm 0.4\times10^2$       & $  6.0\times10^2  \pm 0.4\times10^2$ \\
	$\mu_2$ 		& [keVee]						& $ 44.9\pm 0.4$ 		                          & $46.0 \pm 0.3$ \\
	$\sigma_2$ 	& [keVee]						& $ 3.9 \pm 0.4$  		                          & $ 4.0\pm 0.3$ \\
	
	\hline	 
	\end{tabular}	
	\caption{Fit parameters of the cathode energy spectra obtained after the ER/NR application and a cut on the MPD value.  }
	\label{tab:FitSpectreMPD}
\end{table}

In conclusion, using the MIMAC observables, it is possible to estimate the track position of the main background event sources inside the detector.

\section{Conclusions}

In this paper, we have presented the first results of the analysis of the MIMAC data runs at the LSM.
We have shown, for the first time, the observation of low energy NR 3D tracks from daughter nuclei of the \Nucl{Rn}{222} decay chain.
Finally, using a new MIMAC observable called MPD, we have shown that it is possible to separate anode events from volume and cathode events. We have used this observable in order to show the NR ionization energy spectra. 
This measurement shows the capability of the MIMAC detector and opens the possibility to explore the low energy recoil directionality signature.

Even if the RPR measurement is a validation of the MIMAC detection strategy, it remains a background for dark matter directional detection. 
The next step will be the discrimination of this background using the MIMAC observables,  the coincidence between the chambers and their directionality. This analysis will be described in a dedicated future paper.

The radon gas emanation monitoring and its reduction are major topics for rare event experiments such as dark matter or double-$\beta$ decay search experiments. The new degrees of freedom, offered by the observation of 3D low energy NR tracks describing these events shed a new light on them improving their localization and discrimination.

\section*{Acknowledgements}
We thank the Laboratoire Souterrain de Modane (CNRS/CEA) and its team for their support all along our operation. 
We acknowledge F.~Mayet, J.~Lamblin, J.~Brunner, C.~Couturier and A.~Naver-Agasson for many helpful discussions, and O.~M\'eplan and M.~Ramdhane for radioactivity material measurements performed at the LBA.
MIMAC collaboration acknowledges the ANR-07-BLAN-0255-03 funding.

\bibliography{Progeny}

\providecommand{\href}[2]{#2}\begingroup\raggedright\begin{thebibliography}{10}

\bibitem{PlanckCollaboration2015a}
P.~A.~R. Ade, N.~Aghanim, M.~Arnaud, M.~Ashdown, J.~Aumont, C.~Baccigalupi
  et~al., \emph{{Planck 2015 results}},
  \href{http://dx.doi.org/10.1051/0004-6361/201525830}{\emph{Astronomy {\&}
  Astrophysics} {\bf 594} (oct, 2016) A13},
  [\href{http://arxiv.org/abs/1502.01589}{{\tt 1502.01589}}].

\bibitem{Goodman_1984dc}
M.~W. Goodman and E.~Witten, \emph{{Detectability of Certain Dark Matter
  Candidates}}, \href{http://dx.doi.org/10.1103/PhysRevD.31.3059}{\emph{Phys.
  Rev.} {\bf D31} (1985) 3059}.

\bibitem{Agnes2016}
P.~Agnes, L.~Agostino, I.~F.~M. Albuquerque, T.~Alexander, A.~K. Alton,
  K.~Arisaka et~al., \emph{{Results from the first use of low radioactivity
  argon in a dark matter search}},
  \href{http://dx.doi.org/10.1103/PhysRevD.93.081101}{\emph{Physical Review D -
  Particles, Fields, Gravitation and Cosmology} {\bf 93} (2016) 1--7},
  [\href{http://arxiv.org/abs/1510.00702}{{\tt 1510.00702}}].

\bibitem{Akerib2016a}
D.~S. Akerib, H.~M. Ara{\'{u}}jo, X.~Bai, A.~J. Bailey, J.~Balajthy,
  P.~Beltrame et~al., \emph{Improved limits on scattering of weakly interacting
  massive particles from reanalysis of 2013 lux data},
  \href{http://dx.doi.org/10.1103/PhysRevLett.116.161301}{\emph{Physical Review
  Letters} {\bf 116} (apr, 2016) 161301},
  [\href{http://arxiv.org/abs/1512.03506}{{\tt 1512.03506}}].

\bibitem{XENON100Collaboration2016a}
{XENON100 Collaboration}, E.~Aprile, J.~Aalbers, F.~Agostini, M.~Alfonsi, F.~D.
  Amaro et~al., \emph{{XENON100 Dark Matter Results from a Combination of 477
  Live Days}},  \href{http://arxiv.org/abs/1609.06154}{{\tt 1609.06154}}.

\bibitem{Billard:2013qya}
J.~Billard, L.~Strigari and E.~Figueroa-Feliciano, \emph{Implication of
  neutrino backgrounds on the reach of next generation dark matter direct
  detection experiments},
  \href{http://dx.doi.org/10.1103/PhysRevD.89.023524}{\emph{Phys.Rev.} {\bf
  D89} (2014) 023524}, [\href{http://arxiv.org/abs/1307.5458}{{\tt
  1307.5458}}].

\bibitem{Spergel1988}
D.~Spergel, \emph{{Motion of the Earth and the detection of weakly interacting
  massive particles}},
  \href{http://dx.doi.org/10.1103/PhysRevD.37.1353}{\emph{Phys. Rev. D} {\bf
  37} (Mar., 1988) 1353--1355}.

\bibitem{Billard:2009mf}
J.~Billard, F.~Mayet, J.~Macias-Perez and D.~Santos, \emph{{Directional
  detection as a strategy to discover galactic Dark Matter}},
  \href{http://dx.doi.org/10.1016/j.physletb.2010.06.024}{\emph{Phys.Lett.}
  {\bf B691} (2010) 156--162}, [\href{http://arxiv.org/abs/0911.4086}{{\tt
  0911.4086}}].

\bibitem{Mei:2005gm}
D.~Mei and A.~Hime, \emph{{Muon-induced background study for underground
  laboratories}},
  \href{http://dx.doi.org/10.1103/PhysRevD.73.053004}{\emph{Phys.Rev.} {\bf
  D73} (2006) 053004}, [\href{http://arxiv.org/abs/astro-ph/0512125}{{\tt
  astro-ph/0512125}}].

\bibitem{Bozorgnia:2011vc}
N.~Bozorgnia, G.~B. Gelmini and P.~Gondolo, \emph{{Ring-like features in
  directional dark matter detection}},
  \href{http://dx.doi.org/10.1088/1475-7516/2012/06/037}{\emph{JCAP} {\bf 1206}
  (2012) 037}, [\href{http://arxiv.org/abs/1111.6361}{{\tt 1111.6361}}].

\bibitem{Bozorgnia:2011tk}
N.~Bozorgnia, G.~B. Gelmini and P.~Gondolo, \emph{{Daily modulation due to
  channeling in direct dark matter crystalline detectors}},
  \href{http://dx.doi.org/10.1103/PhysRevD.84.023516}{\emph{Phys.Rev.} {\bf
  D84} (2011) 023516}, [\href{http://arxiv.org/abs/1101.2876}{{\tt
  1101.2876}}].

\bibitem{Billard:2010gp}
J.~Billard, F.~Mayet and D.~Santos, \emph{{Exclusion limits from data of
  directional Dark Matter detectors}},
  \href{http://dx.doi.org/10.1103/PhysRevD.82.055011}{\emph{Phys.Rev.} {\bf
  D82} (2010) 055011}, [\href{http://arxiv.org/abs/1006.3513}{{\tt
  1006.3513}}].

\bibitem{Henderson:2008bn}
S.~Henderson, J.~Monroe and P.~Fisher, \emph{{The Maximum Patch Method for
  Directional Dark Matter Detection}},
  \href{http://dx.doi.org/10.1103/PhysRevD.78.015020}{\emph{Phys.Rev.} {\bf
  D78} (2008) 015020}, [\href{http://arxiv.org/abs/0801.1624}{{\tt
  0801.1624}}].

\bibitem{Billard:2011zj}
J.~Billard, F.~Mayet and D.~Santos, \emph{{Assessing the discovery potential of
  directional detection of Dark Matter}},
  \href{http://dx.doi.org/10.1103/PhysRevD.85.035006}{\emph{Phys.Rev.} {\bf
  D85} (2012) 035006}, [\href{http://arxiv.org/abs/1110.6079}{{\tt
  1110.6079}}].

\bibitem{Green:2010zm}
A.~M. Green and B.~Morgan, \emph{{The median recoil direction as a WIMP
  directional detection signal}},
  \href{http://dx.doi.org/10.1103/PhysRevD.81.061301}{\emph{Phys.Rev.} {\bf
  D81} (2010) 061301}, [\href{http://arxiv.org/abs/1002.2717}{{\tt
  1002.2717}}].

\bibitem{Billard2010}
J.~Billard, F.~Mayet and D.~Santos, \emph{{Markov chain Monte Carlo analysis to
  constrain dark matter properties with directional detection}},
  \href{http://dx.doi.org/10.1103/PhysRevD.83.075002}{\emph{Phys. Rev. D} {\bf
  83} (Apr., 2011) 075002}, [\href{http://arxiv.org/abs/1012.3960}{{\tt
  1012.3960}}].

\bibitem{Alves:2012ay}
D.~S. Alves, S.~E. Hedri and J.~G. Wacker, \emph{{Dark Matter in 3D}},
  \href{http://arxiv.org/abs/1204.5487}{{\tt 1204.5487}}.

\bibitem{Lee:2014cpa}
S.~K. Lee, \emph{{Harmonics in the Dark-Matter Sky: Directional Detection in
  the Fourier-Bessel Basis}},
  \href{http://dx.doi.org/10.1088/1475-7516/2014/03/047}{\emph{JCAP} {\bf 1403}
  (2014) 047}, [\href{http://arxiv.org/abs/1401.6179}{{\tt 1401.6179}}].

\bibitem{O'Hare:2014oxa}
C.~A.~J. O'Hare and A.~M. Green, \emph{{Directional detection of dark matter
  streams}},
  \href{http://dx.doi.org/10.1103/PhysRevD.90.123511}{\emph{Phys.Rev.} {\bf
  D90} (2014) 123511}, [\href{http://arxiv.org/abs/1410.2749}{{\tt
  1410.2749}}].

\bibitem{Grothaus:2014hja}
P.~Grothaus, M.~Fairbairn and J.~Monroe, \emph{{Directional Dark Matter
  Detection Beyond the Neutrino Bound}},
  \href{http://dx.doi.org/10.1103/PhysRevD.90.055018}{\emph{Phys.Rev.} {\bf
  D90} (2014) 055018}, [\href{http://arxiv.org/abs/1406.5047}{{\tt
  1406.5047}}].

\bibitem{Ruppin:2014bra}
F.~Ruppin, J.~Billard, E.~Figueroa-Feliciano and L.~Strigari,
  \emph{{Complementarity of dark matter detectors in light of the neutrino
  background}},
  \href{http://dx.doi.org/10.1103/PhysRevD.90.083510}{\emph{Phys.Rev.} {\bf
  D90} (2014) 083510}, [\href{http://arxiv.org/abs/1408.3581}{{\tt
  1408.3581}}].

\bibitem{Santos2007}
D.~Santos, O.~Guillaudin, T.~Lamy et~al., \emph{{MIMAC: A Micro-TPC Matrix of
  Chambers for direct detection of Wimps}},
  \href{http://dx.doi.org/10.1088/1742-6596/65/1/012012}{\emph{J. Phys. Conf.
  Ser.} {\bf 65} (Apr., 2007) 012012},
  [\href{http://arxiv.org/abs/0703310v1}{{\tt 0703310v1}}].

\bibitem{Ahlen2009}
S.~Ahlen, N.~Afshordi, J.~Battat et~al., \emph{{The case for a directional dark
  matter detector and the status of current experimental efforts}},
  \href{http://dx.doi.org/10.1142/S0217751X10048172}{\emph{Int.J.Mod.Phys.}
  {\bf A25} (2010) 1--51}, [\href{http://arxiv.org/abs/0911.0323}{{\tt
  0911.0323}}].

\bibitem{Battat_2016pap}
J.~B.~R. Battat et~al., \emph{{Readout technologies for directional WIMP Dark
  Matter detection}},
  \href{http://dx.doi.org/10.1016/j.physrep.2016.10.001}{\emph{Phys. Rept.}
  {\bf 662} (2016) 1--46}, [\href{http://arxiv.org/abs/1610.02396}{{\tt
  1610.02396}}].

\bibitem{Santos:2013oua}
D.~Santos, J.~Billard, G.~Bosson et~al., \emph{{MIMAC : A micro-tpc matrix for
  dark matter directional detection}},
  \href{http://dx.doi.org/10.1088/1742-6596/460/1/012007}{\emph{J. Phys. Conf.
  Ser.} {\bf 460} (Oct., 2013) 012007},
  [\href{http://arxiv.org/abs/1304.2255}{{\tt 1304.2255}}].

\bibitem{Iguaz2011}
F.~J. Iguaz, D.~Atti\'{e}, D.~Calvet et~al., \emph{Micromegas detector
  developments for dark matter directional detection with mimac},
  \href{http://dx.doi.org/10.1088/1748-0221/6/07/P07002}{\emph{J. Instrum.}
  {\bf 6} (July, 2011) P07002--P07002},
  [\href{http://arxiv.org/abs/1105.2056}{{\tt 1105.2056}}].

\bibitem{Giomataris2006}
I.~Giomataris, R.~{De Oliveira}, S.~Andriamonje et~al., \emph{{Micromegas in a
  bulk}},
  \href{http://dx.doi.org/10.1016/j.nima.2005.12.222}{\emph{Nucl.Instrum.Meth.}
  {\bf A560} (May, 2006) 405--408}.

\bibitem{Billard2013a}
J.~Billard, F.~Mayet, G.~Bosson et~al., \emph{In situ measurement of the
  electron drift velocity for upcoming directional dark matter detectors},
  \href{http://dx.doi.org/10.1088/1748-0221/9/01/P01013}{\emph{J. Instrum.}
  {\bf 9} (Jan., 2014) P01013--P01013},
  [\href{http://arxiv.org/abs/1305.2360}{{\tt 1305.2360}}].

\bibitem{Richer2009}
J.~Richer, G.~Bosson, O.~Bourrion et~al., \emph{{Development of a front end
  ASIC for Dark Matter directional detection with MIMAC}},
  \href{http://dx.doi.org/10.1016/j.nima.2010.04.041}{\emph{Nucl.Instrum.Meth.}
  {\bf A620} (2010) 470--476}, [\href{http://arxiv.org/abs/0912.0186}{{\tt
  0912.0186}}].

\bibitem{Bourrion2010a}
O.~Bourrion, G.~Bosson, C.~Grignon et~al., \emph{Data acquisition electronics
  and reconstruction software for directional detection of dark matter with
  mimac},
  \href{http://dx.doi.org/10.1016/j.nima.2010.07.035}{\emph{Nucl.Instrum.Meth.}
  {\bf A662} (2010) 207--214}, [\href{http://arxiv.org/abs/1006.1335}{{\tt
  1006.1335}}].

\bibitem{Burgos2007}
S.~Burgos, J.~Forbes, C.~Ghag et~al., \emph{Track reconstruction and
  performance of drift directional dark matter detectors using alpha
  particles},
  \href{http://dx.doi.org/10.1016/j.nima.2007.10.013}{\emph{Nucl.Instrum.Meth.}
  {\bf A584} (Jan., 2008) 114--128},
  [\href{http://arxiv.org/abs/0707.1758}{{\tt 0707.1758}}].

\bibitem{Daw:2013waa}
J.~Brack, E.~Daw, A.~Dorofeev et~al., \emph{{Long-term study of backgrounds in
  the DRIFT-II directional dark matter experiment}},
  \href{http://dx.doi.org/10.1088/1748-0221/9/07/P07021}{\emph{J. Instrum.}
  {\bf 9} (July, 2014) P07021--P07021},
  [\href{http://arxiv.org/abs/1307.5525}{{\tt 1307.5525}}].

\bibitem{Battat2014}
J.~B.~R. Battat, J.~Brack, E.~Daw et~al., \emph{Radon in the drift-ii
  directional dark matter tpc: emanation, detection and mitigation},
  \href{http://arxiv.org/abs/1407.3938}{{\tt 1407.3938}}.

\bibitem{Malling2013}
D.~C. Malling, S.~Fiorucci, M.~Pangilinan et~al., \emph{{Dark Matter Search
  Backgrounds from Primordial Radionuclide Chain Disequilibrium}},
  \href{http://dx.doi.org/10.1016/j.astropartphys.2014.07.009}{\emph{Astropart.
  Phys.} {\bf 62} (May, 2013) 21}, [\href{http://arxiv.org/abs/1305.5183}{{\tt
  1305.5183}}].

\bibitem{ZieglerJ.F.;Biersack1985}
J.~Ziegler and J.~Biersack, \emph{{SRIM - The Stopping and Range of Ions in
  Matter}}.
\newblock Pergamon Press New York, www.srim.org, 1985.

\bibitem{Guillaudin2012}
O.~Guillaudin, J.~Billard, G.~Bosson et~al., \emph{{Quenching factor
  measurement in low pressure gas detector for directional dark matter
  search}}, \href{http://dx.doi.org/10.1051/eas/1253015}{\emph{EAS Publ. Ser.}
  {\bf 53} (Feb., 2012) 119--127}, [\href{http://arxiv.org/abs/1110.2042}{{\tt
  1110.2042}}].

\bibitem{Beringer2012}
J.~Beringer, J.~F. Arguin, R.~M. Barnett et~al., \emph{{Review of Particle
  Physics}}, \href{http://dx.doi.org/10.1103/PhysRevD.86.010001}{\emph{Phys.
  Rev. D} {\bf 86} (July, 2012) 010001}.

\bibitem{Riffard2016mgw}
Q.~Riffard et~al., \emph{{MIMAC low energy electron-recoil discrimination
  measured with fast neutrons}},
  \href{http://dx.doi.org/10.1088/1748-0221/11/08/P08011}{\emph{JINST} {\bf 11}
  (2016) P08011}, [\href{http://arxiv.org/abs/1602.01738}{{\tt 1602.01738}}].

\bibitem{couturier_2017_cathode}
C.~Couturier, D.~Santos, Q.~Riffard et~al.{\emph{To be published in JINST}
  (2007) }.

\bibitem{Biagi1999}
S.~Biagi, \emph{Monte carlo simulation of electron drift and diffusion in
  counting gases under the influence of electric and magnetic fields},
  \href{http://dx.doi.org/10.1016/S0168-9002(98)01233-9}{\emph{Nucl.Instrum.Meth.}
  {\bf A421} (Jan., 1999) 234--240}.

\end{thebibliography}\endgroup


\providecommand{\href}[2]{#2}\begingroup\raggedright\endgroup

\end{document}